\documentclass[11pt]{article}
\usepackage[T1]{fontenc}
\usepackage[utf8]{inputenc}
\usepackage{lmodern}
\usepackage[margin=1in]{geometry}
\usepackage{amsmath,amssymb,amsthm,mathtools,bm}
\usepackage{booktabs,array,longtable}
\usepackage{microtype}
\usepackage{xcolor,xurl,hyperref}
\usepackage{enumitem,listings}
\hypersetup{colorlinks=true,linkcolor=black,citecolor=black,urlcolor=blue!50!black,
 pdftitle={A note on exact external information complexity},
 pdfauthor={Henning Wunderlich}}
\numberwithin{equation}{section}
\newtheorem{theorem}{Theorem}[section]
\newtheorem{proposition}[theorem]{Proposition}
\newtheorem{lemma}[theorem]{Lemma}
\newtheorem{corollary}[theorem]{Corollary}
\theoremstyle{definition}
\newtheorem{definition}[theorem]{Definition}
\newtheorem{remark}[theorem]{Remark}
\newcommand{\AND}{\mathsf{AND}}
\newcommand{\IC}{\operatorname{IC}}
\newcommand{\ICrnd}{B}
\newcommand{\ICdet}{Z}
\newcommand{\ICrndmu}[1]{b_{#1}}
\newcommand{\ICdetmu}[1]{z_{#1}}
\newcommand{\ICcost}[1]{\IC^{\mathrm{ext},#1}_0}
\newcommand{\RS}{\operatorname{RS}_0}
\newcommand{\RSmu}[1]{\operatorname{RS}^{#1}_0}
\newcommand{\supp}{\operatorname{supp}}
\newcommand{\HH}{\mathcal H}
\newcommand{\one}{\mathbf1}
\newcommand{\diag}{\operatorname{diag}}
\newcommand{\E}{\mathbb E}
\newcommand{\R}{\mathbb R}

\newcommand{\KLnat}{D_{\mathrm{nat}}}
\newcommand{\Fmat}{\begin{pmatrix}0&1&1\\1&1&1\\1&1&0\end{pmatrix}}
\newcommand{\Wmat}{\begin{pmatrix}0&1&1\\1&0&1\\1&1&0\end{pmatrix}}
\newcommand{\Jnum}{63460111063493390777838640891505464697531}
\newcommand{\Jden}{67734982792059356713900951995000000000000}
\newcommand{\Lstar}{L_*}
\newcommand{\Ustar}{U_*}

\newcommand{\zenodopath}[1]{\footnote{Computational certificate archive~\cite{WunderlichZenodo2026}, path relative to its root: \nolinkurl{#1}.}}
\title{A note on exact external information complexity}
\author{Henning Wunderlich\thanks{Offenbach am Main. E-mail: \href{mailto:henningwunderlich@t-online.de}{henningwunderlich@t-online.de}.}}
\date{\today}
\begin{document}
\maketitle
\begin{abstract}
We study the prior-free zero-error private-coin external information complexity
$B(f)$ of finite two-party functions, with correctness on every input.
An elementary rectangle/entropy argument gives $B(\AND)=\log_2 3$.
A complete classification of Boolean $3\times3$ matrices leaves, up to
input permutations and output complementation, only the two-overlapping-rectangles
function $F$ without an exact value. An entropy potential and a finite rational
protocol give
$2.1212661667060486842\ldots\le B(F)<2.130998318$.
We also determine its deterministic external information and zero-error
communication costs. A separate product-prior certificate and
internal-information additivity yield
$R_0^{\mathrm{pub}}(F^{\times t})\ge2.021991187178562\ldots\,t$,
strictly exceeding the fractional-cover rate $2t$.
\end{abstract}

\section{Basic notions and notation}
\label{sec:model}
Let $f:\mathcal X\times\mathcal Y\to\mathcal Z$ have finite input alphabets.
Alice receives $x\in\mathcal X$ and Bob receives $y\in\mathcal Y$.
We denote a protocol by $\Pi$ and its transcript on the joint input $XY$ by
$\Pi(XY)$. In a randomized protocol, this transcript also depends on the
parties' private coins; the coins themselves are not part of the transcript.
A protocol specifies the next speaker from the messages already sent and
exchanges ordinary bits. Finite-alphabet messages are encoded by prefix-free
binary strings; there is no communication through silent timing.

A protocol is \emph{fully correct} if it terminates almost surely on every
input and its common terminal output is always $f(x,y)$. Terminal transcripts
are labeled by this output. Let $\mathcal P^{\mathrm{det}}_0(f)$ be the fully
correct deterministic protocols and $\mathcal P^{\mathrm{rnd}}_0(f)$ the fully
correct \emph{private-coin} randomized protocols. The private random tapes are
independent of each other and of the input; no shared random tape is assumed.

\subsection{Deterministic and nondeterministic communication}
The \emph{deterministic communication complexity} is
\begin{equation}
\label{eq:def-D}
D(f)=\min_{\Pi\in\mathcal P^{\mathrm{det}}_0(f)}
       \max_{(x,y)\in\mathcal X\times\mathcal Y}|\Pi(xy)|.
\end{equation}
A \emph{rectangle} is a set $A\times B\subseteq\mathcal X\times\mathcal Y$.
Every correct deterministic protocol partitions the input matrix into
monochromatic leaf rectangles, although not every monochromatic rectangle
partition is induced by a protocol tree.

For Boolean $f$ and $b\in\{0,1\}$, let $C_b(f)$ be the minimum number of
$b$-monochromatic rectangles covering $f^{-1}(b)$. If $f^{-1}(b)$ is
empty, set $C_b(f)=N^b(f)=0$: the verifier always rejects and no
communication or certificate index is needed. For $f^{-1}(b)\ne\varnothing$,
a certificate in the standard nondeterministic model names such a rectangle
and the parties verify membership. With verification bits charged,
\[
 \lceil\log_2 C_b(f)\rceil\le N^b(f)
 \le \lceil\log_2 C_b(f)\rceil+2;
\]
if only the shared rectangle index is charged, the value is
$\lceil\log_2 C_b(f)\rceil$. We write
$N(f)=\max\{N^0(f),N^1(f)\}$ and
$C_{\rm all}(f)=C_0(f)+C_1(f)$. See
\cite[Chapters~1--2]{KN1997} for these standard conventions and their
rectangle-cover formulation.

\subsection{The rectangle-size bound}
\label{sec:rectangle-def}
For a prior $\mu$ on $\mathcal X\times\mathcal Y$, define
\begin{equation}
\label{eq:def-RS}
\rho_\mu(f)=\max_{Q\text{ monochromatic rectangle}}\mu(Q),
\qquad
\RSmu{\mu}(f)=\log_2\frac1{\rho_\mu(f)},
\qquad
\RS(f)=\max_\mu\RSmu{\mu}(f).
\end{equation}
This is the standard distributional rectangle-size bound
\cite[Sections~1.3 and~2.4]{KN1997}. Proposition~\ref{prop:rectangle}
shows directly that the same coefficient-one lower bound applies to
zero-error private-coin external information complexity.

\subsection{External information: protocol, fixed prior, and worst prior}
For either kind of protocol and $XY\sim\mu$, its \emph{external information cost}
is
\begin{equation}
\label{eq:def-protocol-info}
\ICcost{\mu}(\Pi)=I(XY:\Pi(XY)),\qquad XY\sim\mu.
\end{equation}
The colon denotes mutual information. Equivalently, this is
$H_\mu(XY)-H_\mu(XY\mid\Pi(XY))$. It measures what an observer of the communicated
transcript learns about the joint input \cite{BGPW2015,CoverThomas2006}.
For deterministic $\Pi$, the transcript is a function of $XY$, and
\begin{equation}
\label{eq:det-protocol-cost}
\ICcost{\mu}(\Pi)=H_\mu(\Pi(XY)),
\end{equation}
the entropy of its protocol partition.

We use the abbreviations
\[
 B(f):=\IC^{\mathrm{ext},\mathrm{prv}}_0(f),\qquad
 Z(f):=\IC^{\mathrm{ext},\mathrm{det}}_0(f),
\]
for randomized private-coin and deterministic external information complexity,
respectively. Both are defined with the prior fixed before the protocol is
optimized. Their fixed-prior versions are denoted by $b_\mu(f)$ and
$z_\mu(f)$.

\begin{definition}[Deterministic external information complexity]
\label{def:quantities}
For a fixed prior and then for the worst prior, respectively, set
\begin{align}
\ICdetmu{\mu}(f)
 &=\min_{\Pi\in\mathcal P^{\mathrm{det}}_0(f)}\ICcost{\mu}(\Pi),
 \label{eq:def-detmu}\\
\ICdet(f)
 &=\max_\mu\ICdetmu{\mu}(f).
 \label{eq:def-det}
\end{align}
The prior is fixed \emph{before} the protocol is optimized.
\end{definition}

\begin{definition}[Randomized external information complexity]
For private-coin randomized protocols, analogously, set
\begin{align}
\ICrndmu{\mu}(f)
 &=\inf_{\Pi\in\mathcal P^{\mathrm{rnd}}_0(f)}\ICcost{\mu}(\Pi),
 \label{eq:def-rndmu}\\
\ICrnd(f)
 &=\max_\mu\ICrndmu{\mu}(f).
 \label{eq:def-rnd}
\end{align}
The infimum allows nonattainment by an ordinary protocol; when attained it
is a minimum. This qualification permits limiting optimal protocols
\cite[Section~3.2]{FHLY2019}.
\end{definition}

In both definitions the subscript $0$ requires zero error on \emph{all} inputs,
including those of zero prior mass. The protocol may depend on $\mu$; correctness
may not. The outer maxima are well-defined on the compact simplex of priors:
for a fixed protocol its information cost is continuous, and an infimum of
these continuous functions is upper semicontinuous. The deterministic minimum
is attained because only finitely many protocol partitions need be considered.
A single protocol that works well for every prior supplies an upper bound,
not the definition.

The randomized quantity also has the protocol-first representation
\begin{equation}
\label{eq:randomized-minimax}
 B(f)=\inf_{\Pi\in\mathcal P^{\mathrm{rnd}}_0(f)}
       \sup_\mu I_\mu(XY;\Pi(XY)).
\end{equation}
This is the external-information minimax identity
\cite[Theorem~3.17]{BravermanIC}. An input-independent choice between
correct protocols can be sampled privately and announced; revealing the
choice averages their information costs. In contrast, $Z(f)$ is not
identified with the protocol-first deterministic quantity
$\min_\Pi\max_\mu H_\mu(\Pi(XY))=\log_2 C^P(f)$, where $C^P(f)$ is
the minimum number of reachable protocol leaves.

All randomization in the defining classes is private. The finite symmetry
seed used later is generated privately and transmitted; see
Section~\ref{sec:private-symmetry}. Public randomness may also be used as
an analytical convenience when it is included in the observer's transcript;
this does not change the information infima \cite{BravermanIC}.
Public random bits are free only in the separate public-coin
communication-cost measure.

Information, ordinary entropy $H$, and relative entropy $D$ are in bits.
When natural logarithms are used, we write $\ln$, $\HH$, or
$D_{\mathrm{nat}}$ explicitly, with $0\log0=0$.

\section{Results and related work}
\label{sec:intro}
Our two examples are the one-bit AND function and the matrix
\[
\AND=\begin{pmatrix}0&0\\0&1\end{pmatrix},\qquad
F=\Fmat.
\]
The seven one-inputs of $F$ form the union of two monochromatic rectangles
that meet at $(2,2)$, hence the name \emph{two-overlapping-rectangles
function}.  Equivalently, $F$ is disjointness on a two-element universe
when each input set is empty or a singleton.  This is, up to row and column permutations and output complementation,
the unique $3\times3$ Boolean matrix for which the elementary
rectangle-size lower bound and a minimum-leaf deterministic protocol
do not match (Theorem~\ref{thm:classification}).  It also separates
rectangle-size, randomized external information, deterministic external
information, and zero-error communication under the conventions of
Section~\ref{sec:model}.

For AND we reproduce a short argument based on the chain
\[
\RS(f)\le\ICrnd(f)\le\ICdet(f),
\]
with equality at $\log_2 3$.  The author established $B(\AND)=\log_2 3$ during his doctoral work
twenty years ago, using an argument based on deterministic external
information complexity. That proof was not published. The external value
was subsequently published by Braverman, Garg, Pankratov, and Weinstein
\cite{BGPW2013,BGPW2015}; the distributional and internal-information
analyses in that work are distinct from the short prior-free argument here.

The classification of all $3\times3$ Boolean matrices is given in
Section~\ref{sec:classification}.  The author investigated that
classification about twenty years before the present write-up; like the
simplified AND argument, it was not published at the time.

Our principal quantitative result concerns the exceptional function $F$:
\begin{theorem}[Computer-assisted bounds]\label{thm:main}
Under the distribution-first private-coin, common-output zero-error
conventions of Section~\ref{sec:model},
\begin{equation}\label{eq:maininterval}
\Lstar:=2.1212661667060486842\ldots\le\ICrnd(F)
<\Ustar:=2.130998318.
\end{equation}
The lower endpoint has the exact expression in
\eqref{eq:fullhard} and \eqref{eq:J-corrected}.  A separately concave
entropy potential and an explicit finite protocol provide the two
computer-assisted certificates.
\end{theorem}
The certificates verify separate properties: a global entropy-majorization
inequality for all permitted signals, and correctness together with an
input-wise relative-entropy bound for one protocol. The authoritative data
and verification programs supporting these two endpoints are deposited in
a public reproducibility archive~\cite{WunderlichZenodo2026}; their scope
and arithmetic assumptions are summarized in Appendix~\ref{app:repro}.
The exact value of $\ICrnd(F)$ is unknown.

\begin{proposition}[Other complexities of $F$]\label{prop:overview}
With the conventions in Section~\ref{sec:model},
\[
\RS(F)=2,\qquad
\ICdet(F)=2.2690572336533368\ldots,\qquad D(F)=3.
\]
The deterministic value has an exact scalar characterization in
Appendix~\ref{app:det}.  The expected zero-error communication costs are
$R_0^{\mathrm{pub}}(F)=5/2$ and $R_0^{\mathrm{priv}}(F)=3$.
\end{proposition}

\begin{corollary}[Unique $3\times3$ information-complexity separation]
\label{cor:unique-gap}
Up to independent row and column permutations and output
complementation, $F$ is the only $3\times3$ Boolean function with
\[\ICrnd(f)<\ICdet(f).\]
\end{corollary}
\begin{proof}
Theorem~\ref{thm:classification} gives equality for every other
class.  For $F$, Theorem~\ref{thm:main} and
Proposition~\ref{prop:overview} give the strict inequality.
\end{proof}

\subsection{Related literature}
The standard rectangle and cover framework is presented in Kushilevitz and
Nisan \cite{KN1997}. Dietzfelbinger and Wunderlich study the corresponding
protocol-partition entropy and average-case formulation \cite{DW2007}.  Rectangle and information lower-bound techniques also appear
in the work of Bar-Yossef, Jayram, Kumar, and Sivakumar
\cite{BJKS2004}, and Kerenidis, Laplante, Lerays, Roland, and Xiao
\cite{KLLRX2015}.  The latter's bounded-error, predominantly internal
information results should not be cited as the coefficient-one
zero-error \emph{external} rectangle inequality used here; we prove that
inequality directly in Section~\ref{sec:and}.

Braverman et al.\ established the exact external value of AND and a
local characterization of zero-error information complexity
\cite{BGPW2013,BGPW2015}.  Filmus, Hatami, Li, and You subsequently
refined the AND analysis, including a correction to a boundary argument
and a distinction between all-input and support-only correctness
\cite{FHLY2019}.  We use only the sufficient full-signal direction of the
external concealed-entropy framework.  Ma and Ishwar studied related
concave-envelope and infinite-message methods for interactive source
coding under a different operational model \cite{MaIshwar2013}.

The broader context includes the information/amortized-communication
results of Braverman and Rao \cite{BravermanRao2014}, computability
of information complexity \cite{BravermanSchneider2016}, and separations
involving external information \cite{GanorKolRaz2016,BravermanMinzer2021}.
The zero-error landscape is particularly sensitive to communication-cost
conventions and distributions without full support
\cite{KMSY2016,DFHL2018,Li2022}.  Suruga's 2026 preprint
establishes an amortized expected-communication characterization in terms
of a zero-error \emph{internal} information quantity
\cite{Suruga2026}; it does not evaluate our external-information measure.
The one-copy numerical bounds do not rely on an asymptotic direct-sum
theorem. Section~\ref{sec:direct-sums} separately applies the
external-information minimax identity and internal-information additivity
\cite{BravermanIC} to a product-prior certificate.

\subsection{Contribution and limitations}
Beyond the AND proof, the contributions are a classification of all
$3\times3$ Boolean communication matrices that isolates $F$ as the
unique exceptional class, a certified interval for its randomized external
information, an exact deterministic comparison, and an explicit method
for checking external-information bounds through
concealed-entropy potentials and input-wise transcript divergence.
Appendix~\ref{app:pricing} develops the supporting-plane dual, the
three-outcome reduction, and the graph curvature and relative-entropy
identities underlying local continuation pricing. Appendix~\ref{app:additional}
relates local choices to a common worst-prior optimization and gives
opposite-player curvature and signal-compatibility criteria.
Appendix~\ref{sec:prefix} gives general conditional potential bounds at a
transcript cut and reach-weighted improvement budgets.
The numerical endpoints in Theorem~\ref{thm:main} are computer-assisted and
use the finite certificate package deposited with this paper. These
certificates are not proof-assistant formalizations; Appendix~\ref{app:repro}
states the arithmetic assumptions and the scope of the independent verifiers.

\paragraph{Repeated evaluation.}
Set
\[
 p=(9/20,1/10,9/20),\qquad
 \beta=2H(p)-179/250=2.021991187178562\ldots.
\]
The product prior $p\otimes p$ admits a separate entropy certificate,
proved in Appendix~\ref{app:product-prior}. Section~\ref{sec:direct-sums}
uses internal-information additivity to obtain, for every $t\ge1$,
\[
 R_0^{\mathrm{pub}}(F^{\times t})\ge B(F^{\times t})\ge t\beta>2t.
\]
Here all $t$ output values must be computed, and communication is measured
by worst-input expected length. The fractional-cover benchmark is exactly
$2t$. The improvement uses a product-posterior constraint, not a general
direct-sum theorem for prior-free external information.

\paragraph{Organization.}
Section~\ref{sec:and} gives the rectangle inequality and the AND proof.
Section~\ref{sec:classification} classifies all Boolean $3\times3$
matrices. The subsequent sections compare the information and communication
measures for $F$, establish the three-class reduction, summarize the
one-copy certificates, and derive the repeated-task bounds.
The appendices provide the exact deterministic calculation, the lower and
upper constructions, the continuation-pricing analysis, the product-prior
certificate, and the reproducibility specifications.

\section{The rectangle inequality and the exact external value of AND}
\label{sec:and}
\subsection{The coefficient-one rectangle inequality}
Recall the rectangle quantities $\rho_\mu(f)$, $\RSmu{\mu}(f)$, and $\RS(f)$ from Section~\ref{sec:rectangle-def}. For a private-coin protocol $\Pi$, the probability of any transcript $t$ factors as
\begin{equation}
\label{eq:factorization}
\Pr[\Pi(XY)=t\mid X=x,Y=y]=a_t(x)b_t(y).
\end{equation}
This follows by multiplying Alice's and Bob's successive conditional message probabilities. The actual likelihood support $Q_t=\{a_t>0\}\times\{b_t>0\}$ is a rectangle. Full correctness makes a terminal $Q_t$ monochromatic. For a correlated prior, the posterior support itself need not be a rectangle: it is only contained in $Q_t$.

\begin{proposition}[Rectangle size versus external information]
\label{prop:rectangle}
For every fully correct protocol $\Pi$ and every prior $\mu$,
\begin{equation}
\label{eq:rectangle-lower}
\ICcost{\mu}(\Pi)\ge\log_2\frac1{\rho_\mu(f)}.
\end{equation}
Consequently, for every prior,
\[
\RSmu{\mu}(f)\le\ICrndmu{\mu}(f)\le\ICdetmu{\mu}(f),
\]
and in the prior-free setting,
\begin{equation}
\label{eq:chain}
\RS(f)\le\ICrnd(f)\le\ICdet(f).
\end{equation}
\end{proposition}
\begin{proof}
Let $\mu_t$ be the posterior at a terminal transcript of positive probability. Since $\mu_t$ is supported on $Q_t$,
\begin{align*}
D(\mu_t\Vert\mu)
 &=D\bigl(\mu_t\Vert\mu(\,\cdot\mid Q_t)\bigr)
   +\log_2\frac1{\mu(Q_t)}\\
 &\ge\log_2\frac1{\rho_\mu(f)}.
\end{align*}
Average over transcripts, using the posterior-divergence identity $\ICcost{\mu}(\Pi)=\E_tD(\mu_t\Vert\mu)$. This proves \eqref{eq:rectangle-lower}; the identity is standard \cite{CoverThomas2006}. For each fixed prior take the infimum over private-coin protocols, and then maximize over priors. This gives the first inequality in \eqref{eq:chain}. The second follows because deterministic protocols are a subclass and their transcripts are functions of the input.
\end{proof}

For a deterministic partition $\mathcal Q_\Pi$, the same lower bound is the elementary estimate
\[
H_\mu(\Pi(XY))=\sum_{Q\in\mathcal Q_\Pi}\mu(Q)\log_2\frac1{\mu(Q)}
\ge\log_2\frac1{\rho_\mu(f)}.
\]
This is the deterministic entropy form of the standard rectangle bound
\cite{KN1997}; compare also the average-case formulation in \cite{DW2007}.
The posterior proof shows directly why private randomness does not invalidate
its coefficient-one external counterpart.

\begin{lemma}[Leaf entropy]
\label{lem:leaf}
If a deterministic protocol $\Pi$ has $k$ nonempty leaves, then
\[\max_\mu\ICcost{\mu}(\Pi)=\log_2 k.\]
Hence, if $\ell(f)$ is the minimum number of nonempty leaves in a
correct deterministic protocol, $\ICdet(f)\le\log_2\ell(f)$;
equality is not asserted in general.
\end{lemma}
\begin{proof}
The transcript entropy is at most $\log_2 k$. Equality for this fixed protocol
is attained by putting mass $1/k$ on one input from each leaf. Choosing a
minimum-leaf protocol bounds the inner minimum for every prior. Nonempty
leaves of a correct deterministic protocol on finite alphabets are finite in
number after unreachable nodes are removed.
\end{proof}

\subsection{The AND benchmark and an unpublished proof}
The phrase ``one-bit AND'' refers to one input bit held by each party. Some cited papers call the same function ``two-bit AND.''

\begin{theorem}[Published exact value; rectangle/entropy proof]
\label{thm:and}
For $\AND(x,y)=x\land y$,
\[
\RS(\AND)=\ICrnd(\AND)=\ICdet(\AND)=\log_2 3.
\]
\end{theorem}
\begin{proof}
Put equal mass on $(0,1),(1,0),(1,1)$ and zero mass on $(0,0)$:
\[
\mu_* =\begin{pmatrix}0&1/3\\1/3&1/3\end{pmatrix}.
\]
No monochromatic rectangle contains two of these three inputs. The two zero inputs $(0,1)$ and $(1,0)$ cannot share a rectangle because their rectangular hull contains $(1,1)$; that input has the other output. Thus $\rho_{\mu_*}(\AND)=1/3$. Proposition~\ref{prop:rectangle} proves the lower bound for every randomized protocol. Equivalently, the transcript identifies the supported input, so its external information is $H_{\mu_*}(XY)=\log_2 3$.

For the upper bound, Alice sends $x$. If $x=0$, the parties output zero; otherwise Bob sends $y$ and both output $y$. The three terminal rectangles have probabilities
\[
p_0=\mu(0,0)+\mu(0,1),\qquad p_{10}=\mu(1,0),\qquad p_{11}=\mu(1,1).
\]
For \emph{every} prior, the information cost is the partition entropy $H(p_0,p_{10},p_{11})$. Convexity of $\varphi(t)=t\log_2t$ gives
\[
\frac13\sum_i\varphi(p_i)\ge\varphi(1/3),
\qquad
H(p_0,p_{10},p_{11})\le\log_2 3.
\]
Equality holds under $\mu_*$. The pointwise rectangle lower bound and the single deterministic upper protocol therefore match after optimizing protocols for each prior and then maximizing over priors.
\end{proof}

\begin{remark}[Attribution and scope]
The author established $B(\AND)=\log_2 3$ during his doctoral work
twenty years ago using deterministic external information complexity.
That proof was not published. The published external value is BGPW's
Theorem~2.3, restated and proved as Theorem~7.3 in their expanded manuscript
\cite{BGPW2013,BGPW2015}. Their proof also uses the three-point prior and
three-transcript entropy maximum. The formulation here makes the
coefficient-one rectangle inequality explicit for both deterministic and
randomized protocols. It does not invoke the more elaborate
distribution-dependent buzzer analysis, and it does not assert that
randomization fails to help AND under every fixed prior.
\end{remark}

\section{Why the two-overlapping-rectangles function is exceptional}
\label{sec:classification}
We now classify the $3\times3$ Boolean functions under independent
permutations of rows and columns.  Both information complexities and
\(\RS\) are also invariant under transposition (exchanging the players)
and complementation (interchanging the output labels).  We use these
additional symmetries to state the classification compactly.

Call $S\subseteq\mathcal X\times\mathcal Y$ \emph{rectangle-separated}
if every monochromatic rectangle meets $S$ in at most one input.
Unlike an ordinary one-color fooling set, $S$ may contain inputs of
both output values.  Let $\ell(f)$ be the minimum number of leaves of
a deterministic common-output protocol for $f$.

\begin{lemma}[Matching rectangle and leaf bounds]
\label{lem:classification-sandwich}
If $f$ has a rectangle-separated set of size $k$ and a correct
deterministic protocol with $k$ leaves, then
\[
\RS(f)=\ICrnd(f)=\ICdet(f)=\log_2 k.
\]
\end{lemma}
\begin{proof}
Give the $k$ selected inputs equal probability.  Every monochromatic
rectangle has probability at most $1/k$, hence $\RS(f)\ge\log_2 k$.
Proposition~\ref{prop:rectangle} gives the same lower bound on randomized
external information complexity.  Conversely, a deterministic protocol
with $k$ leaves has partition entropy at most $\log_2 k$ under any prior,
so Lemma~\ref{lem:leaf} gives $\ICdet(f)\le\log_2 k$.  These bounds
also apply in the distribution-first convention.
\end{proof}

\begin{theorem}[Classification of $3\times3$ Boolean matrices]
\label{thm:classification}
There are $36$ equivalence classes of $3\times3$ Boolean matrices under
independent row and column permutations.  Of these, $24$ contain a
repeated row or column; the remaining $12$ reduce to five types when
transposition and output complementation are also allowed.

In $34$ of the $36$ classes, the maximum size of a rectangle-separated
set equals the minimum number of deterministic protocol leaves.  The
only exceptions are $F$ and $1-F$, up to row and column permutations:
for each of them the two numbers are $4$ and $5$, respectively.
Consequently, every other class satisfies
\[
\RS(f)=\ICrnd(f)=\ICdet(f)
\in\{0,1,\log_2 3,2,\log_2 5,\log_2 6\}.
\]
\end{theorem}

\begin{proof}
Burnside's lemma gives the row/column orbit count
\[
\frac{512+6\cdot64+4\cdot8+9\cdot32+12\cdot4+4\cdot8}{36}=36.
\]
The $36$ classes form the $13$ families in
Table~\ref{tab:classification}.  The table gives a rectangle-separated
set of size $k$ and a correct protocol with $L$ leaves, with $k=L$ in
all but the indicated exceptional family.  Their explicit inputs and
protocol trees, together with an exhaustive independent enumeration,
are given in Appendix~\ref{app:classification}.  For $F$ and its
complement, no rectangle-separated set has five elements and no
protocol has four leaves; this also follows directly from the
combinatorial argument in Proposition~\ref{prop:combinatorial} and the
pairwise rectangle condition.  Lemma~\ref{lem:classification-sandwich}
therefore applies to the other $34$ classes.
\end{proof}

\begin{table}[htbp]
\centering\small
\caption{The $13$ families under row and column permutations,
transposition, and output complementation.  A representative is written
as its three binary rows.  The multiplicity $m$ counts classes under
\emph{row and column permutations alone}.  Here $k$ is the maximum
rectangle-separated set size and $L$ the minimum protocol leaf count.}
\label{tab:classification}
\begin{tabular}{@{}lrrrl@{}}
\toprule
Representative & $m$ & $k$ & $L$ & Structure\\
\midrule
\multicolumn{5}{@{}l}{\emph{A repeated row or column: $24$ classes}}\\
\texttt{000/000/000} &2&1&1&constant\\
\texttt{000/000/001} &2&3&3&smaller alphabet\\
\texttt{000/000/011} &4&3&3&smaller alphabet\\
\texttt{000/000/111} &4&2&2&one-sided\\
\texttt{000/001/110} &4&4&4&smaller alphabet\\
\texttt{000/001/111} &4&4&4&smaller alphabet\\
\texttt{000/011/011} &2&3&3&smaller alphabet\\
\texttt{001/001/110} &2&4&4&smaller alphabet\\
\midrule
\multicolumn{5}{@{}l}{\emph{Three different rows and columns: $12$ classes}}\\
\texttt{000/001/010} &2&4&5&$F$ or $1-F$\\
\texttt{000/001/011} &2&5&5&path $P_4$ and isolates\\
\texttt{001/010/100} &2&6&6&permutation identity\\
\texttt{000/011/101} &4&5&5&path $P_5$ and isolate\\
\texttt{001/010/101} &2&6&6&$P_4\sqcup K_2$\\
\bottomrule
\end{tabular}
\end{table}

The bipartite-graph descriptions in the last three rows use the
$1$-entries of the displayed representatives.  Thus, besides the
permutation identity and $F$, there are three further irreducible
patterns; their information complexities nevertheless follow from the
elementary rectangle/leaf sandwich.  More explicitly, among the
$36$ row/column classes, the numbers with common exact value
$0$, $1$, $\log_2 3$, $2$, $\log_2 5$, and $\log_2 6$ are
$2,4,8,10,6$, and $4$, respectively; the remaining two are the
exceptional pair.

\begin{remark}[Historical note]
\label{rem:old-classification}
The author investigated this small-matrix classification approximately
two decades before the present write-up but did not publish the
observation.  The classification and its explicit witnesses are recorded
here to explain why $F$ is the natural first exceptional example, rather
than as a claim that there are no other $3\times3$ combinatorial types.
\end{remark}

\section{The exceptional two-overlapping-rectangles function}
\label{sec:F}
Rows and columns of $F$ are indexed by $1,2,3$. Its maximal two-row one-rectangles are
\[
R_+=\{1,2\}\times\{2,3\},\qquad
R_-=\{2,3\}\times\{1,2\},
\qquad R_+\cap R_-=\{(2,2)\}.
\]
The middle row and middle column are additional maximal one-rectangles. Each zero cell is necessarily a singleton monochromatic rectangle.

\subsection{Covers, partitions, and deterministic information}
\begin{proposition}
\label{prop:combinatorial}
The minimum cover numbers are $C_0(F)=C_1(F)=2$. The minimum number $C_{\rm all}(F)$ of monochromatic rectangles of either color covering all inputs is four. Moreover,
\[
\RS(F)=2,\qquad \ell(F)=5,\qquad D(F)=3.
\]
\end{proposition}
\begin{proof}
The two zero singletons and $R_+,R_-$ give the asserted covers. Neither color can be covered by a single monochromatic rectangle. The four corner inputs
\[
\{(1,1),(1,3),(3,1),(3,3)\}
\]
form a fooling set for rectangles of either color: no monochromatic rectangle contains two corners. The uniform corner prior gives $\RS(F)\ge2$. Conversely, for any prior one of the four covering rectangles has mass at least $1/4$, so $\RS(F)\le2$.

Any monochromatic partition has two separate zero parts. The seven one-inputs cannot be partitioned into two rectangles. In a putative such partition, $(1,3)$ and $(3,1)$ belong to different rectangles. The rectangle containing $(1,3)$ must also contain $(1,2)$ and $(2,3)$, since neither can share a one-rectangle with $(3,1)$. It therefore contains $(2,2)$. Symmetrically the other rectangle also contains $(2,2)$, a contradiction. At least five parts are necessary.

A five-leaf protocol exists: Alice first tests whether $x=2$. On that branch both output one. Otherwise she identifies which endpoint row she holds, and Bob sends whether the input is the zero cell in that row. This protocol has depth at most three. Lemma~\ref{lem:leaf} gives the upper bound $\ICdet(F)\le\log_2 5$, not an equality for the distribution-first definition. A depth-two binary tree has at most four leaves, so $D(F)=3$.
\end{proof}

For deterministic external information complexity as defined in \eqref{eq:def-det}, Appendix~\ref{app:det} proves
\begin{equation}
\label{eq:det-value}
\ICdet(F)=\log_2(2+K_*)=2.2690572336533368416\ldots,
\end{equation}
where $K_*$ is specified by a unique scalar root. Its proof enumerates the coarsest protocol partitions and then optimizes their entropies by concavity. In particular, $\ICdet(F)<\log_2 5$: the five-leaf upper bound is not tight when the protocol is optimized separately for each prior.

\subsection{Expected zero-error communication}
Write
\[
R_0^{\rm pub}(f)=\inf_\Pi\max_{x,y}\E[|\Pi(xy)|],
\]
with public and private randomness permitted, and define $R_0^{\rm priv}$ by excluding public randomness. Both costs require almost-sure correct termination on every input. For communication cost, unlike external information, revealing previously private tapes can be used in the standard public-coin simulation without charging information leakage.

\begin{proposition}
\label{prop:R0}
Under these expected-cost, common-output conventions,
\[
R_0^{\mathrm{pub}}(F)=\frac52,
\qquad R_0^{\mathrm{priv}}(F)=3.
\]
\end{proposition}
\begin{proof}
Consider the deterministic protocol in which Alice sends whether $x=1$. If yes, Bob sends whether $y=1$ and the computation ends. If not, Bob sends whether $y=3$. A negative answer ends the computation with output one; on a positive answer Alice sends whether $x=3$. Its length matrix is
\[
L_1=\begin{pmatrix}2&2&2\\2&2&3\\2&2&3\end{pmatrix}.
\]
Simultaneously exchanging $1$ and $3$ gives a protocol with
\[
L_3=\begin{pmatrix}3&2&2\\3&2&2\\2&2&2\end{pmatrix}.
\]
A fair public choice between them has expected length at most $5/2$ on every input.

Each zero input requires at least two bits in any deterministic protocol: one bit cannot isolate both a row and a column. The two zeros cannot both terminate at depth two. For example, if Alice starts, the first-bit branch leading to a two-bit zero transcript at $(1,1)$ must isolate row~1, while the branch for $(3,3)$ must isolate row~3. There is then no first-bit branch available on row~2. The argument is symmetric when Bob starts. Hence
\[
|\Pi(11)|+|\Pi(33)|\ge5.
\]
Fix all random tapes of a randomized protocol. Since the input set is finite, almost every tape choice gives a correct terminating deterministic protocol on all inputs. Averaging the last inequality yields $R_0^{\rm pub}(F)\ge5/2$.

For a private-coin protocol, the first speaker is fixed. Suppose Alice starts. A positive-probability two-bit transcript outputting zero at $(1,1)$ requires the likelihood support of its first bit to be the singleton row~1: the second bit must constrain Bob's column and cannot remove another row. Similarly, such a transcript at $(3,3)$ requires a first-bit support consisting only of row~3. Both first-bit values cannot have these supports, because Alice must send some bit on input~2. Thus at least one zero input has no positive-probability transcript of length at most two. Its expected length is at least three. Symmetry treats a Bob start. The deterministic depth-three protocol gives the reverse bound.
\end{proof}

\subsection{Which separation is being asserted?}
The numerical interval in Theorem~\ref{thm:main}, together with the preceding calculations, gives
\begin{equation}
\label{eq:strictchain}
2<\ICrnd(F)<\ICdet(F)<R_0^{\rm pub}(F)<D(F).
\end{equation}
For clarity, the values are collected in Table~\ref{tab:measures}.
\begin{table}[htbp]
\centering
\caption{Values under the conventions of this paper. Information and expected communication are in bits.}
\label{tab:measures}
\begin{tabular}{@{}ll@{}}
\toprule
Quantity & Value for $F$\\
\midrule
Both-color logarithmic cover number $\log_2C_{\rm all}$ & $2$\\
Rectangle-size bound $\RS$ & $2$\\
Randomized external information $\ICrnd$ & $[\,2.121266166706048\ldots,\ 2.130998318\,)$\\
Deterministic external information $\ICdet$ & $2.269057233653336\ldots$\\
Public-coin expected zero-error communication & $5/2$\\
Private-coin expected zero-error communication & $3$\\
Deterministic communication & $3$\\
\bottomrule
\end{tabular}
\end{table}

Under the certificate-only convention described in Section~\ref{sec:model},
the one-color quantities $\lceil\log_2 C_0(F)\rceil$ and
$\lceil\log_2 C_1(F)\rceil$ are both one, while
$\log_2 C_{\rm all}(F)=2$. These cover-based values must not be silently
substituted for the bit-counting $N(F)$, which includes communicated verification
bits. The strict chain therefore uses the explicitly defined rectangle bound
rather than $N$. Likewise, replacing public by private coins destroys the strict
inequality between $R_0$ and $D$ for this matrix.

\section{An exact reduction and uniform channel bounds}
\label{sec:reduction}
Let $S_1=F^{-1}(1)$; it contains seven inputs, including the center. Write
\begin{equation}
\label{eq:reduced-definition}
\kappa=
\max_{\nu:\supp\nu\subseteq S_1}\ICrndmu{\nu}(F).
\end{equation}
This is an abbreviation for a restricted maximization of the already defined
fixed-prior complexity, not a different order of optimization. The protocols
inside $\ICrndmu{\nu}(F)$ must still be correct on all nine inputs.
For a \emph{fixed} fully correct private-coin protocol $\Pi$, write
\[
k_\Pi=\max_{\nu:\supp\nu\subseteq S_1}\ICcost{\nu}(\Pi)
\]
for its reduced channel capacity. We keep a fixed channel's capacity distinct
from the information complexity of a function.

\begin{proposition}[Three-class identities]
\label{prop:threeclass}
For a fixed fully correct private-coin protocol,
\begin{equation}
\label{eq:threeclass-fixed}
\max_\mu\ICcost{\mu}(\Pi)=\log_2(2+2^{k_\Pi}).
\end{equation}
With the distribution-first definition \eqref{eq:def-rnd},
\begin{equation}
\label{eq:threeclass}
\ICrnd(F)=\log_2(2+2^\kappa).
\end{equation}
\end{proposition}
\begin{proof}
A zero-output terminal rectangle must be either $\{(1,1)\}$ or $\{(3,3)\}$.
Thus every correct transcript distinguishes the three classes
$\{(1,1)\}$, $\{(3,3)\}$, and $S_1$. For
\[
\mu=p\delta_{11}+q\delta_{33}+s\nu,
\qquad p+q+s=1,\quad \supp\nu\subseteq S_1,
\]
the chain rule gives, for each fixed protocol,
\begin{equation}
\label{eq:threeclass-mi}
\ICcost{\mu}(\Pi)=H(p,q,s)+s\ICcost{\nu}(\Pi).
\end{equation}
There is no conditional information within either singleton zero class.
For a fixed channel, maximize first over $\nu$ and then over $(p,q,s)$.
The maximum of $H(p,q,s)+sk$ is $\log_2(2+2^k)$, attained at
\[
(p,q,s)=\left(\frac1{2+2^k},\frac1{2+2^k},\frac{2^k}{2+2^k}\right).
\]
This proves \eqref{eq:threeclass-fixed}.

For the function complexity, instead \emph{first take the infimum over
protocols} in \eqref{eq:threeclass-mi} for this fixed $\mu$. The fully correct
private-coin protocol class is the same for $\mu$ and $\nu$, so
\begin{equation}
\label{eq:threeclass-distributional}
\ICrndmu{\mu}(F)=H(p,q,s)+s\ICrndmu{\nu}(F).
\end{equation}
For $s=0$ the last term is zero and the identity still holds. Maximizing this
identity over $\nu$ and $(p,q,s)$ proves \eqref{eq:threeclass}.
No minimax interchange is needed. The same distributional argument applies
with $\mathrm{det}$ in place of $\mathrm{rnd}$.
\end{proof}

The reduction removes the optimization over the total masses of the two zero
inputs. It does not remove the center from $S_1$ or show that a maximizing
reduced prior assigns it zero mass.

\subsection{Information radius and a worst-prior criterion}
Let $W_z$ be a fixed protocol's transcript law on input $z$, and let $Q$
dominate all the relevant laws. For any prior $\mu$, set
$Q_\mu=\sum_z\mu_zW_z$. The chain rule for relative entropy gives the
extended-real identity
\[
\sum_{z:\mu_z>0}\mu_zD(W_z\Vert Q)
=\ICcost{\mu}(\Pi)+D(Q_\mu\Vert Q).
\]
When $D(W_z\Vert Q)<\infty$ for every relevant input $z$, this can be
rearranged as
\begin{equation}
\label{eq:radius}
\ICcost{\mu}(\Pi)
=\sum_z\mu_zD(W_z\Vert Q)-D(Q_\mu\Vert Q)
\le\max_zD(W_z\Vert Q).
\end{equation}
The final inequality provides a uniform information upper bound and follows
from nonnegativity of relative entropy \cite{CoverThomas2006}. The same upper
inequality follows from the non-subtracted identity when some divergences
are infinite, without forming an undefined difference of infinities. For
our finite protocols, domination by $Q$ makes all these divergences finite.

For a finite channel, $I_\mu$ is concave in $\mu$. Its maximizing-prior conditions are
\begin{equation}
\label{eq:capacity-kkt}
D(W_z\Vert Q_{\mu_*})\le K,
\qquad
\mu_*(z)>0\ \Longrightarrow\ D(W_z\Vert Q_{\mu_*})=K.
\end{equation}
They are also sufficient by \eqref{eq:radius}. At an interior coordinate, differentiating $I_\mu$ gives $D(W_z\Vert Q_\mu)-1/\ln2$, and the common constant disappears under $\sum_z\mu_z=1$. Boundary coordinates use one-sided derivatives. This is a channel-capacity instance of the usual Karush--Kuhn--Tucker conditions \cite{BoydVandenberghe2004}.

A rational reference prior used in a certificate need not satisfy \eqref{eq:capacity-kkt} exactly. Its role is only to produce a dominating $Q$ with a small largest divergence. Near equality of several input divergences is evidence about that fixed channel's capacity, not proof of optimality among protocols.

\subsection{Symmetry with a communicated private seed}
\label{sec:private-symmetry}
The four-element group generated by transposition and simultaneous exchange
of labels $1$ and $3$ preserves $F$. Alice privately selects a uniform group
element $S$, transmits its two-bit encoding, and the parties run the relabeled
protocol using fresh private coins. No public random tape is needed. If $M$
is the later message string, independence of $S$ and $XY$ gives
\begin{equation}
\label{eq:seed-cost}
I(XY:(S,M))=I(XY:M\mid S).
\end{equation}
Thus the information cost is the average of the relabeled costs, and its
largest cost over priors cannot exceed that of the original protocol.
Generating and communicating input-independent randomness in this way is
standard for information complexity \cite[Section~2.3]{FHLY2019}.
The seed must remain in the transcript. Sending it adds two communication
bits but no information about the input; this is not an identification of
public- and private-coin communication complexity.

A symmetric fixed channel has a symmetric capacity-maximizing prior: average
any maximizing prior over the group and use concavity of mutual information
in the prior. Such a prior has the form
\[
\begin{pmatrix}a&b&c\\b&d&b\\c&b&a\end{pmatrix}.
\]
The distributional function $\mu\mapsto\ICrndmu{\mu}(F)$ is also concave,
being the infimum of concave fixed-protocol costs, and is invariant under the
same group. Thus a worst prior for the distribution-first complexity may
likewise be chosen symmetric. The one-input reference prior in
Section~\ref{sec:upper} has $a=d=0$, but zero center mass remains a chosen
certificate parameter, not a theorem about the unrestricted optimum.

\section{The two computer-assisted estimates}
\label{sec:bounds}
We give the constructions and the deductions of the bounds here.  The
exact coefficient arrays, terminal checks, and protocol reconstruction
are specified in Appendices~\ref{sec:lower}, \ref{sec:upper}, and
\ref{app:repro}; no empirical optimizer-convergence assertion is used as
a proof.

\subsection{Lower bound: a concave concealed-entropy potential}
For nonnegative weights $W=(W_{ij})$, write
\[
\HH(w)=\Bigl(\sum_z w_z\Bigr)\ln\Bigl(\sum_z w_z\Bigr)
             -\sum_z w_z\ln w_z,
\qquad E_W(r,c)=\HH((W_{ij}r_ic_j)_{ij}).
\]
The likelihood vectors $r,c$ encode the effect of a transcript on the
input distribution.  A separately concave, separately homogeneous
function $U$ that lies between $0$ and $E_W$, and agrees with $E_W$ on
all terminal monochromatic rectangles, satisfies
\begin{equation}\label{eq:overview-potential}
I_\nu(XY:\Pi(XY))\ge
\frac{\HH(W)-U(\one,\one)}{(\sum_{ij}W_{ij})\ln2},
\qquad \nu=W/\sum W_{ij},
\end{equation}
for every fully correct private-coin protocol $\Pi$.  Indeed, each
local signal splits one likelihood vector into a sum of child vectors;
separate concavity bounds the sum of child potentials by the parent.
The inequalities telescope to terminal conditional entropy.  The exact
argument is Lemma~\ref{lem:potential}.

We take the hard reduced prior
\begin{equation}\label{eq:prior-summary}
\nu_*=\frac{1}{148}
\begin{pmatrix}0&25&24\\25&0&25\\24&25&0\end{pmatrix}.
\end{equation}
Its zero center mass is a choice of prior, \emph{not} an input promise:
valid protocols must compute $F$ correctly on the center as well.
The certificate starts with a separately degree-eight rational Bernstein
polynomial $P_8$, and subtracts three compatible, piecewise
rational--logarithmic terms.  Their supports lie strictly inside the
likelihood simplices, so the terminal conditions remain unchanged.
A global curvature verification establishes separate concavity of the
combined supporting function.  With the homogeneous extension and
entropy cap, \eqref{eq:overview-potential} gives
\[
\ICrndmu{\nu_*}(F)\ge1.2331249058620917801\ldots.
\]
The closed-form starting value is stated in
\eqref{eq:J-corrected}; the full-prior three-class reduction yields
\begin{equation}\label{eq:lower-main}
\ICrnd(F)\ge\Lstar
=\log_2\!\left(2+\frac{148}{25}
\left(\frac{25}{24}\right)^{12/37}e^{-J_*}\right),
\end{equation}
where $J_*$ is given exactly in the appendix.

\subsection{Upper bound: one finite protocol and its information radius}
The upper certificate supplies a private-coin protocol, represented by
rational local signal rules and fixed continuation protocols.  The
protocol is correct on \emph{all nine} inputs.  A finite symmetry
choice is privately sampled and then transmitted, so the construction
does not require free public randomness.

For input $z=(x,y)$ let $W_z$ be its conditional transcript law.
For any reference transcript law $Q$ dominating them, the
relative-entropy identity
\begin{equation}\label{eq:radius-summary}
I_\mu(XY:\Pi(XY))
=\sum_z\mu(z)D(W_z\Vert Q)
-D\!\left(\sum_z\mu(z)W_z\middle\Vert Q\right)
\le\max_zD(W_z\Vert Q)
\end{equation}
holds for every prior $\mu$.  For the seven one-inputs, the verified
strict divergence bounds, in bits, are
\begin{center}
\begin{tabular}{@{}lr@{}}\toprule
input class & $D(W_z\Vert Q)<$\\\midrule
adjacent $12,21,23,32$ & $1.251085696967958$\\
cross $13,31$ & $1.251085696967991$\\
center $22$ & $0.216212811303620$\\\bottomrule
\end{tabular}
\end{center}
The reference law $Q$ is the output law at a rational six-point
reference prior.  The center divergence is checked despite the zero
reference input mass.  The independent proof of correctness and of
these numerical inequalities is specified in Appendix~\ref{sec:upper}.

By the three-class identity \eqref{eq:threeclass-fixed},
\[
\ICrnd(F)\le\sup_\mu I_\mu(XY:\Pi(XY))
<\log_2(2+2^{1.251085696967991})<2.130998318.
\]
This finishes the proof of Theorem~\ref{thm:main}.

\section{Direct-sum bounds for zero-error communication}
\label{sec:direct-sums}

We combine the one-copy estimates of Theorem~\ref{thm:main} with
internal-information additivity to obtain lower bounds for repeated evaluation
of the two-overlapping-rectangles function. A product-prior certificate yields
a lower bound of more than $2.02$ bits per copy for both external
information and expected communication, strictly exceeding the fractional-cover
rate of two bits. The argument applies more generally to
external-information lower bounds under priors with limited input correlation.
All logarithms and information quantities in this section are in bits.

\subsection{Repeated tasks and communication}
\label{subsec:ds-definitions}
We use $b_\mu(h)$ and $B(h)$ from Section~\ref{sec:model}, with the
protocol-first representation \eqref{eq:randomized-minimax}.
In this section, $T$ includes any public randomness in addition to the
communicated messages; private random tapes are not included.
The quantity $R_0^{\rm pub}(h)$ is the worst-input expected communication
of a fully correct public-coin protocol. Since external information is
bounded by expected communication,
\begin{equation}
 B(h)\le R_0^{\rm pub}(h).
 \label{eq:ds-info-comm}
\end{equation}
For $t\ge1$, the repeated task is
\[
 h^{\times t}(x_1,\ldots,x_t;y_1,\ldots,y_t)
   =\bigl(h(x_1,y_1),\ldots,h(x_t,y_t)\bigr).
\]
The protocol must produce every output, not their conjunction or XOR.
Independent execution gives
\begin{equation}
 B(h^{\times t})\le tB(h).
 \label{eq:ds-upper}
\end{equation}
Indeed, the capacity of a product of independent transcript channels is
additive. This upper bound therefore holds even though the supremum in
$B(h^{\times t})$ ranges over arbitrary joint input distributions.

\subsection{Fractional-cover lower bounds}
\label{subsec:ds-cover}
Let \(\operatorname{fcov}_{\pm}(h)\) denote the minimum total weight of a
fractional cover of the entire communication matrix by monochromatic
rectangles. Fractional-cover duality identifies the rectangle-size
quantity from \eqref{eq:def-RS} as
\[
 \RS(h)=\log_2\operatorname{fcov}_{\pm}(h).
\]
For Boolean \(h\), the cover number inside the logarithm is the sum of the two
one-sided fractional cover numbers. This convention also applies to
vector-valued outputs.

Fractional-cover duality yields an input distribution under which every
monochromatic rectangle has mass at most \(2^{-\RS(h)}\). The
positive-probability support of a zero-error private-coin transcript is a
monochromatic rectangle. Consequently, the posterior divergence from this
input distribution is at least \(\RS(h)\) for every transcript, and
averaging proves \(B(h)\ge\RS(h)\). Fractional covers are multiplicative
under the vector-output product \cite[Theorem~2.10]{KKN}; hence
\begin{equation}
 R_0^{\rm pub}(h^{\times t})\ge B(h^{\times t})\ge t\RS(h).
 \label{eq:ds-cover-rate}
\end{equation}

For total Boolean functions on \(n\)-bit inputs, fractional-cover rounding and
the Aho--Ullman--Yannakakis simulation \cite{KKN,AUY} imply that, for universal
constants \(a,b>0\),
\begin{equation}
 R_0^{\rm pub}(h^{\times t})
 \ge t\bigl[a\sqrt{R_0^{\rm pub}(h)}-b\log_2(n+2)\bigr]_+,
 \qquad [u]_+:=\max\{u,0\}.
 \label{eq:ds-sqrt}
\end{equation}
This bound follows from the fractional-cover rate alone. A direct-sum lower
bound proportional to \(B(h)\) would be stronger when \(B(h)\) substantially
exceeds \(\RS(h)\).

\subsection{Correlation-penalized information bounds}
\label{subsec:ds-K}
For a protocol with transcript \(T\) under \(\mu\), write
\[
 e_\mu:=I_\mu(XY;T),\qquad
 i_\mu:=I_\mu(X;T\mid Y)+I_\mu(Y;T\mid X),\qquad
 j_\mu:=I_\mu(X;Y)-I_\mu(X;Y\mid T).
\]
The chain rule and the interactive transcript structure give
\begin{equation}
 e_\mu=i_\mu+j_\mu,\qquad 0\le j_\mu\le I_\mu(X;Y).
 \label{eq:ds-eij}
\end{equation}
The nonnegativity of \(j_\mu\) is specific to communication transcripts; for
product inputs, \(j_\mu=0\) and \(e_\mu=i_\mu\)
\cite[Proposition~3.12]{BravermanIC}. Define
\[
 \mathcal I_\mu(h):=\inf_{\pi\in\mathcal P^{\mathrm{rnd}}_0(h)}i_\mu(\pi),
 \qquad
 K(h):=\sup_\mu\bigl[b_\mu(h)-I_\mu(X;Y)\bigr].
\]
Taking infima in \eqref{eq:ds-eij} yields
\begin{equation}
 \mathcal I_\mu(h)\le b_\mu(h)
 \le\mathcal I_\mu(h)+I_\mu(X;Y).
 \label{eq:ds-i-b}
\end{equation}
Internal information complexity is additive under independent repetition,
including under the all-input zero-error convention
\cite[Theorem~4.2]{BravermanIC}:
\begin{equation}
 \mathcal I_{\mu^{\otimes t}}(h^{\times t})=t\mathcal I_\mu(h).
 \label{eq:ds-internal-add}
\end{equation}
Independence is required between copies, not between the two inputs within
each copy.

\begin{theorem}[Correlation-penalized direct-sum bound]
\label{thm:ds-K}
For every finite function \(h\), input distribution \(\mu\), and integer
\(t\ge1\),
\begin{equation}
 R_0^{\rm pub}(h^{\times t})\ge B(h^{\times t})
 \ge t\bigl[b_\mu(h)-I_\mu(X;Y)\bigr].
 \label{eq:ds-pointwise-rate}
\end{equation}
In particular,
\begin{equation}
 R_0^{\rm pub}(h^{\times t})\ge B(h^{\times t})\ge tK(h).
 \label{eq:ds-K-rate}
\end{equation}
\end{theorem}
\begin{proof}
Equations~\eqref{eq:ds-i-b} and \eqref{eq:ds-internal-add} give
\[
 B(h^{\times t})\ge b_{\mu^{\otimes t}}(h^{\times t})
 \ge\mathcal I_{\mu^{\otimes t}}(h^{\times t})
 =t\mathcal I_\mu(h)
 \ge t\bigl[b_\mu(h)-I_\mu(X;Y)\bigr].
\]
Apply \eqref{eq:ds-info-comm} and then take the supremum over \(\mu\).
\end{proof}

Theorem~\ref{thm:ds-K} transfers one-copy external-information lower bounds to
repeated computation through internal-information additivity. In particular,
\begin{equation}
 0\le K(h)\le\sup_\mu\mathcal I_\mu(h).
 \label{eq:ds-K-internal}
\end{equation}
It does not assert additivity of external information under arbitrary priors.

Let \(B_{\rm prod}(h):=\sup_{\mu_X\otimes\mu_Y}
 b_{\mu_X\otimes\mu_Y}(h)\). Then \(B_{\rm prod}(h)\le K(h)\).
For a product prior \(\mu=\mu_X\otimes\mu_Y\), equality of internal and
external information, together with \eqref{eq:ds-internal-add}, gives
\begin{equation}
 b_{\mu^{\otimes t}}(h^{\times t})=t b_\mu(h).
 \label{eq:ds-product-add}
\end{equation}
It follows that \(B_{\rm prod}(h)=B(h)\) is sufficient for exact additivity
\(B(h^{\times t})=tB(h)\). This condition is not assumed for \(F\).

\subsection{External information in the two-copy extraction}
\label{subsec:ds-extraction}
The distinction between internal and external additivity is explicit in the
two-copy reduction. Let \(Z_1=(X_1,Y_1)\sim\mu_1\) and
\(Z_2=(X_2,Y_2)\sim\mu_2\) be independent, and let \(T\) be the transcript
of a protocol for \((f,g)\). Braverman's extraction
\cite[Figure~3]{BravermanIC} publicly samples \(Y_2\) and privately samples
\(X_2\mid Y_2\) to obtain a protocol for \(f\). For \(g\), it publicly
samples \(X_1\) and privately samples \(Y_1\mid X_1\). The resulting external
costs are
\[
 e_f^{(1)}=I(Z_1;T\mid Y_2),\qquad
 e_g^{(1)}=I(Z_2;T\mid X_1).
\]
They satisfy
\begin{equation}
 e_f^{(1)}+e_g^{(1)}
 =I(Z_1Z_2;T)+I(X_1;Y_2\mid T).
 \label{eq:ds-cross-one}
\end{equation}
Indeed, the chain-rule expansion leaves the correction
\(I(X_1;Y_2\mid T)-I(X_2;Y_1\mid X_1,Y_2,T)\). The second term vanishes by
independence of the input pairs and the transcript likelihood factorization.
The extracted internal costs add exactly, so their correlation-loss terms
exceed the joint correlation loss by \(I(X_1;Y_2\mid T)\).

In the reverse orientation, the costs are
\(e_f^{(2)}=I(Z_1;T\mid X_2)\) and
\(e_g^{(2)}=I(Z_2;T\mid Y_1)\), with correction
\(I(Y_1;X_2\mid T)\). Writing
\(\bar e_f=(e_f^{(1)}+e_f^{(2)})/2\) and analogously for \(\bar e_g\),
we obtain
\begin{equation}
 I(Z_1Z_2;T)=\bar e_f+\bar e_g
 -\frac12\bigl[I(X_1;Y_2\mid T)+I(Y_1;X_2\mid T)\bigr].
 \label{eq:ds-cross-average}
\end{equation}
The public padding is included in each extracted transcript. Thus the chain
rule alone does not allocate external information additively: an argument
replacing \(K(h)\) by \(B(h)\) must also control the cross-instance
correlations in \eqref{eq:ds-cross-average}.

\subsection{Correlation of maximizing priors}
\label{subsec:ds-E}
Define
\[
 \mathcal M(h):=\{\mu:b_\mu(h)=B(h)\},\qquad
 E_{\min}(h):=\min_{\mu\in\mathcal M(h)}I_\mu(X;Y).
\]
These extrema are attained for finite input alphabets. To see this, let
\(\delta\) be the total variation distance between two input laws. A
common-mixture decomposition bounds the difference between their mutual
informations, uniformly over transcript channels, by
\(h_2(\delta)+\delta\log_2(|\mathcal X||\mathcal Y|)\). Taking the infimum
over the fixed class of correct protocols preserves this modulus of
continuity. Hence \(b_\mu(h)\) is continuous, \(\mathcal M(h)\) is nonempty
and compact, and the minimum defining \(E_{\min}(h)\) exists.

Choosing a minimum-correlation maximizing prior in
Theorem~\ref{thm:ds-K} gives
\begin{equation}
 R_0^{\rm pub}(h^{\times t})\ge B(h^{\times t})
 \ge t[B(h)-E_{\min}(h)]_+.
 \label{eq:ds-E-rate}
\end{equation}
More generally, for \(\varepsilon\ge0\), put
\begin{equation}
 E_\varepsilon(h):=
 \min_{\mu:\,b_\mu(h)\ge B(h)-\varepsilon}I_\mu(X;Y).
 \label{eq:ds-Eepsilon}
\end{equation}
Then
\begin{equation}
 B(h^{\times t})\ge
 t[B(h)-\varepsilon-E_\varepsilon(h)]_+.
 \label{eq:ds-Eepsilon-rate}
\end{equation}
In particular, a product prior satisfying \(b_\mu(h)\ge\beta\), together with
an upper bound \(B(h)\le U\), certifies
\begin{equation}
 E_{U-\beta}(h)=0,\qquad K(h)\ge\beta.
 \label{eq:ds-near-hard}
\end{equation}
This conclusion requires neither an exact maximizing prior nor the value of
\(E_{\min}(h)\). The inequality
\([B(h)-E_{\min}(h)]_+\le K(h)\) also shows that optimizing the
correlation-penalized rate can be stronger than restricting to exact maximizers.

The correlation penalty cannot in general be bounded by a fixed fraction of
\(B(h)\). Equality on \(n\)-bit strings has \(O(1)\) zero-error internal
information \cite[Proposition~3.22]{BravermanIC}, whereas its external
information is \(\Theta(n)\). By \eqref{eq:ds-i-b}, every maximizing prior
for equality therefore has input mutual information at least \(B(h)-O(1)\).
The criterion \eqref{eq:ds-E-rate} consequently captures only part of the
external-information lower bound in general.

\subsection{Application to the two-overlapping-rectangles function}
\label{subsec:ds-application}
For the two-overlapping-rectangles function, Theorem~\ref{thm:main} gives
\begin{equation}
 F=\begin{pmatrix}0&1&1\\1&1&1\\1&1&0\end{pmatrix},\qquad
 \Lstar\le B(F)<\Ustar,\qquad \Ustar=2.130998318,
 \label{eq:ds-imported-interval}
\end{equation}
where \(\Lstar=2.1212661667060486842\ldots\) is the exact expression in
Theorem~\ref{thm:main}.

The integral and fractional monochromatic cover numbers of \(F\) both equal
four. The two rectangles
\(\{1,2\}\times\{2,3\}\) and \(\{2,3\}\times\{1,2\}\) cover the
one-inputs, and the zero-inputs form two singleton rectangles. Conversely,
every monochromatic rectangle contains at most one of the four corner inputs;
assigning dual weight one to each corner proves the matching fractional lower
bound. Product covers and fractional-cover multiplicativity therefore give
\begin{equation}
 \operatorname{fcov}_{\pm}(F^{\times t})
 =\operatorname{cov}_{\pm}(F^{\times t})=4^t,
 \qquad \RS(F^{\times t})=2t.
 \label{eq:ds-F-cover}
\end{equation}
Thus the benchmark \eqref{eq:ds-cover-rate} is exactly \(2t\), without an
integral-cover rounding loss.

Appendix~\ref{app:product-prior} gives the separate product-prior
entropy certificate:
\begin{equation}
 \begin{gathered}
 p=\left(\frac9{20},\frac1{10},\frac9{20}\right),\qquad \mu_*=p\otimes p,\\
 b_{\mu_*}(F)\ge\beta,\qquad
 \beta:=2H(p)-\frac{179}{250}
       =2.0219911871785624425\ldots.
 \end{gathered}
 \label{eq:ds-product-input}
\end{equation}
This distributional estimate is independent of the prior-free lower bound
\(\Lstar\). The finite inequalities used in this certificate are displayed
and proved for the stated orientation of $F$ in
Appendix~\ref{app:product-prior}; no large numerical data are required.

\begin{corollary}[Improvement over the fractional-cover rate]
\label{cor:ds-F}
For every integer \(t\ge1\),
\begin{equation}
 \max\{\Lstar,t\beta\}\le B(F^{\times t})<t\Ustar,\qquad
 R_0^{\rm pub}(F^{\times t})\ge\max\{\Lstar,t\beta\}.
 \label{eq:ds-F-main}
\end{equation}
In particular,
\begin{equation}
 R_0^{\rm pub}(F^{\times t})\ge
 2.0219911871785624425\ldots\,t.
 \label{eq:ds-F-comm}
\end{equation}
\end{corollary}
\begin{proof}
Since \(I_{\mu_*}(X;Y)=0\), Theorem~\ref{thm:ds-K} and
\eqref{eq:ds-product-input} imply \(B(F^{\times t})\ge t\beta\).
Fixing all but one input pair gives \(B(F^{\times t})\ge B(F)\ge \Lstar\).
The upper bound follows from independent execution and
\eqref{eq:ds-imported-interval}. Finally apply \eqref{eq:ds-info-comm}.
\end{proof}

The coefficient exceeds the exact fractional-cover rate by
\[
 \beta-2=0.0219911871785624425\ldots,
 \qquad \frac{\beta}{2}-1=0.0109955935892812\ldots,
\]
an improvement of approximately \(1.10\%\). This improvement uses the
product-posterior constraint in the one-copy information certificate;
rectangular support alone does not suffice. Indeed,
\[
 \sup_\nu\left[-\log_2\max_{R\text{ monochromatic}}\nu(R)
                     -I_\nu(X;Y)\right]\le\RS(F)=2.
\]
The tensorization step remains the standard internal-information theorem.

Combining the product-prior certificate with the uniform upper bound gives
\begin{equation}
 E_{\varepsilon_*}(F)=0,\qquad
 \varepsilon_*:=\Ustar-\beta=0.1090071308214375575\ldots,
 \label{eq:ds-F-near-hard}
\end{equation}
and
\begin{equation}
 \frac{B(F^{\times t})}{tB(F)}
 \ge\frac{\beta}{B(F)}>
 \frac{\beta}{\Ustar}=0.9488469184134581013\ldots.
 \label{eq:ds-F-retention}
\end{equation}
Thus a product prior attains external-information complexity within
\(0.109008\) bits of \(B(F)\), and repeated evaluation retains more than
\(94.88\%\) of the one-copy prior-free external information per copy.
The corresponding estimate from the fractional-cover bound and the same
upper bound is \(2/\Ustar=0.9385272541543132274\ldots\). A smaller upper bound
\(\Ustar\) would improve \eqref{eq:ds-F-near-hard} and
\eqref{eq:ds-F-retention}, but not the absolute lower-bound coefficient
\(\beta\) in \eqref{eq:ds-F-comm}.

\subsection{Scope of the repeated-task bounds}
\label{subsec:ds-status}
The interval \(\Lstar\le B(F)<\Ustar\) and the distributional estimate
\eqref{eq:ds-product-input} have different roles. The latter proves
\(K(F)\ge\beta\) and hence the unconditional repeated-task lower bound
\(t\beta\). The former gives the information upper bound \(t\Ustar\) and the
one-copy lower bound \(\Lstar\), but does not imply \(K(F)\ge \Lstar\), determine
\(E_{\min}(F)\), or establish
\([B(F)-E_{\min}(F)]_+>2\). In particular, \(B(F^{\times t})<t\Ustar\) is not
an upper bound on communication complexity.

If exact additivity \(B(F^{\times t})=tB(F)\) were established, the one-copy
lower certificate would yield the stronger communication rate
\(\Lstar=2.1212661667060486842\ldots\). More generally, a bound
\(B(F^{\times t})\ge c\,tB(F)\) would imply
\(R_0^{\rm pub}(F^{\times t})\ge c\Lstar t\), exceeding the cover rate whenever
\[
 c>\frac2\Lstar=0.9428331207986249075\ldots.
\]
These conclusions are conditional. The bound in Corollary~\ref{cor:ds-F}
requires neither prior-free additivity nor an exact hardest prior, and can be
improved directly by strengthening the product-prior certificate.

Known failures of external-information direct sum concern distinct
formulations. Kol--Moran--Shpilka--Yehudayoff distinguish support-only
correctness from all-input correctness in \cite[Section~4]{KMSY2016};
Braverman--Minzer also obtain fixed-distribution zero-error separations under
all-input correctness \cite[Corollary~1.4]{BravermanMinzer2021}. These results preclude a
general fixed-prior additivity principle, but do not supply a conclusion about
the supremum over priors defining \(B\). The improvement proved here is a
strict constant improvement over the fractional-cover benchmark for this
fixed function, obtained from a concrete product-prior certificate rather
than a general external-information direct-sum theorem.

\section{Discussion and open problem}
\label{sec:discussion}
The classification identifies $F$ and its complement as the only
$3\times3$ Boolean matrices, up to row and column permutations, for
which the rectangle-size bound and the deterministic leaf upper bound
fail to coincide.  For AND, those bounds already agree at $\log_2 3$.
For the exceptional $3\times3$ function, the rectangle-size bound is strictly weaker than
the certified randomized lower bound, which is in turn below the exact
deterministic information complexity.  The quantitative certificate
gap is less than $0.009733$ bits; the exact randomized value remains
unknown.

Optimality within a finite continuation family does not establish optimality
among all protocols. A stronger family should be closed under the useful
splittings of \emph{both} parties. At each state, a splitting followed by
selected child continuations yields a new valid protocol. For an achievable
concealed-entropy value to be optimal, it must meet both parties' global
concavity inequalities as well as the terminal values.
Appendix~\ref{app:additional} develops criteria for this compatibility, and
Appendix~\ref{sec:prefix} bounds the possible improvement after a fixed opening.

The repeated-task bound in Corollary~\ref{cor:ds-F} uses a different
one-copy witness: a product prior for which external and internal information
coincide. It improves the fractional-cover rate without identifying a
maximizing prior for $B(F)$. The global interval in
Theorem~\ref{thm:main} does not by itself justify multiplying its lower
endpoint by the number of copies. A matching product-prior certificate,
or a stronger direct-sum theorem for $B$, would be an additional result.

A numerical curiosity is that
\[
\log_2\frac{\Gamma(1/4)^4}{4\pi^2}
=2.1299025767961573543\ldots
\]
lies inside the interval of Theorem~\ref{thm:main}.  The gamma
expression occurs as the reciprocal hyperbolic metric density at
$-1$ on $\mathbb C\setminus\{0,1\}$ with curvature normalized to
$-1$ \cite{KimMaMa2015}.  No mathematical connection to information
complexity, or equality with this value, has been established.

The principal outstanding mathematical task is to construct a
matching protocol and globally feasible entropy potential. 

\section*{Disclosure of AI assistance}
OpenAI ChatGPT was used as an assistance tool for drafting, checks, and locating related literature.  
The author reviewed, revised, and takes responsibility for the mathematical statements, arguments, citations, and final contents of the manuscript.

\appendix

\section{Exact deterministic external information complexity}
\label{app:det}
This appendix proves \eqref{eq:det-value} with the definition \eqref{eq:def-det}: for each prior minimize the deterministic protocol's information cost, then maximize over priors.

\subsection{Coarsest protocol partitions}
For fixed $\mu$, refining a deterministic protocol partition cannot reduce its entropy. It therefore suffices to consider partitions that cannot be coarsened to another protocol partition. There are finitely many such partitions on a $3\times3$ input alphabet.

The function
\[
g(\mu):=\ICdetmu{\mu}(F)=\min_{\Pi\in\mathcal P_0^{\mathrm{det}}(F)}\ICcost{\mu}(\Pi)
\]
is concave, since it is the pointwise minimum of finitely many concave entropy functions. It is invariant under the four symmetries of $F$. Averaging a maximizing prior over this group therefore gives a symmetric maximizer
\begin{equation}
\label{eq:detprior}
\mu=\begin{pmatrix}a&b&c\\b&d&b\\c&b&a\end{pmatrix},
\qquad 2a+4b+2c+d=1.
\end{equation}

A complete recursive enumeration gives twelve coarsest protocol partitions. Each contains the zero singleton probabilities $a,a$. The remaining probabilities and multiplicities are shown in Table~\ref{tab:partitions}.
\begin{table}[htbp]
\centering
\caption{The coarsest deterministic protocol partitions under the symmetric prior \eqref{eq:detprior}. The two zero-leaf probabilities $a,a$ are omitted.}
\label{tab:partitions}
\begin{tabular}{@{}clc@{}}
\toprule
Type & One-leaf probabilities & Number\\
\midrule
A & $b,\ b+c,\ c+2b+d$ & 4\\
B & $c,\ c,\ 2b,\ 2b+d$ & 2\\
C & $b+c,\ b+c,\ 2b+d$ & 2\\
E & $c,\ b+c,\ 2b,\ b+d$ & 4\\
\bottomrule
\end{tabular}
\end{table}

For completeness, the enumeration is defined by the following finite recurrence. For each rectangle $A\times B$, if it is monochromatic, retain its one-part partition. Otherwise, for each nontrivial split $A=A_0\dot\cup A_1$, combine every pair of already computed partitions of $A_0\times B$ and $A_1\times B$; do the analogous operation for every nontrivial split of $B$. Remove duplicate partitions and those refining another retained protocol partition. Every deterministic first bit is one of these splits, so induction on the rectangle dimensions proves exhaustiveness. Conversely, each recurrence step is a legal first message and hence produces a protocol partition. Coarsening a subprotocol remains legitimate in its parent tree, so removing refinements loses no entropy minimum.

\subsection{Removing center mass}
Replace $(b,d)$ by $(b+d/4,0)$, keeping $a,c$ fixed. This preserves normalization. In type~A it transfers $d/4$ from the largest one-leaf mass to each of the two smaller masses. In type~B it balances $2b$ and $2b+d$. Both entropies weakly increase by concavity.

At $d=0$, the type~C entropy is at least the type~A entropy: replacing the pair $(b,c+2b)$ by $(b+c,2b)$ balances it. Type~E is obtained from type~A by splitting the part $c+2b$ into $c$ and $2b$, so its entropy is also at least that of type~A. Therefore the entropy minimum at the transformed prior is the smaller of types~A and~B and is no smaller than the old four-type minimum. A maximizing prior can be chosen with $d=0$.

It follows that
\begin{equation}
\label{eq:detoptimization}
\ICdet(F)=\max_{2a+4b+2c=1}
\min\{H(a,a,b,b+c,c+2b),\ H(a,a,c,c,2b,2b)\}.
\end{equation}

\subsection{The scalar root}
For $b>0$ put $r=c/b$. Conditional on the input being a one-input, the two competing entropies are
\begin{align*}
A(r)&=H\left(\frac1{2r+4},\frac{r+1}{2r+4},\frac{r+2}{2r+4}\right)
=1+\frac12 h_2\!\left(\frac1{r+2}\right),\\
B(r)&=H\left(\frac r{2r+4},\frac r{2r+4},\frac2{2r+4},\frac2{2r+4}\right)
=1+h_2\!\left(\frac r{r+2}\right).
\end{align*}
Here $h_2$ is binary entropy in bits. The total zero mass can be optimized independently, exactly as in Proposition~\ref{prop:threeclass}. For fixed $r$ the value is
\[
\log_2\bigl(2+2^{\min\{A(r),B(r)\}}\bigr).
\]
The function $A$ strictly decreases for $r>0$, while $B$ strictly increases on $0<r<2$. They cross once in that interval. Beyond $r=2$, $A(r)$ is already below its crossing value, so no larger optimum occurs there. The endpoint $b=0$ is the corresponding limiting case and also gives a smaller value.

The crossing equation is
\begin{equation}
\label{eq:detroot}
(r+2)\ln(r+2)+(r+1)\ln(r+1)
=2r\ln r+4\ln2.
\end{equation}
The derivative of the left side minus the right side is
$\ln((r+2)(r+1)/r^2)>0$, proving uniqueness on $(0,\infty)$. The sign is negative at $r=0$ by continuity and positive at $r=2$. Denote its root by $r_*$ and put
\begin{equation}
\label{eq:detK}
K_*=(2r_*+4)\exp\left[-\frac{(r_*+2)\ln(r_*+2)+(r_*+1)\ln(r_*+1)}{2r_*+4}\right].
\end{equation}
Then $K_*=2^{A(r_*)}$, and we have proved
\[
\ICdet(F)=\log_2(2+K_*)=2.2690572336533368416\ldots.
\]
The root is $r_*=0.2437094730170131111\ldots$. A maximizing prior in \eqref{eq:detprior} has
\[
a=\frac1{2+K_*},\qquad
b=\frac{K_*}{(2r_*+4)(2+K_*)},\qquad
c=r_*b,\qquad d=0.
\]
The scalar equation and monotonicity prove optimality; the displayed decimal approximations are not used in that proof.

\section{A certified lower bound from concealed entropy}
\label{sec:lower}
\subsection{A full-signal potential inequality}
Let $W=(W_{ij})$ be a nonnegative weight matrix, $S=\sum_{ij}W_{ij}>0$, and $\nu=W/S$. Define homogeneous entropy in \emph{nats} by
\[
\HH(w)=\left(\sum_z w_z\right)\ln\left(\sum_z w_z\right)-\sum_z w_z\ln w_z,
\]
with $\HH(0)=0$. At a likelihood state $(r,c)$ set
\begin{equation}
\label{eq:Edef}
E_W(r,c)=\HH\bigl((W_{ij}r_ic_j)_{ij}\bigr).
\end{equation}
Likelihood factorization makes these precisely the possible unnormalized posterior weights. The actual remaining input rectangle is $\supp(r)\times\supp(c)$, independent of which entries of $W$ vanish.

\begin{lemma}[Concealed-entropy certificate]
\label{lem:potential}
Suppose $U$ on the nonnegative likelihood cones is separately concave, positively homogeneous in each vector, and satisfies
\[
0\le U(r,c)\le E_W(r,c).
\]
Assume also that $U(r,c)=E_W(r,c)$ whenever $\supp(r)\times\supp(c)$ is monochromatic for $f$. Then every fully correct protocol satisfies
\begin{equation}
\label{eq:potential-bound}
\ICcost{\nu}(\Pi)\ge
\frac{\HH(W)-U(\one,\one)}{S\ln2}.
\end{equation}
\end{lemma}
\begin{proof}
If Alice sends a signal with conditional probabilities $p_b(i)$, the children have row likelihoods $r^{(b)}_i=r_ip_b(i)$ and $\sum_b r^{(b)}=r$. Concavity and homogeneity imply superadditivity on the cone, hence
\[
U(r,c)\ge\sum_bU(r^{(b)},c).
\]
Bob's messages give the same inequality in $c$. At a terminal transcript $t$, the equality condition identifies $U$ with $\HH((W_{ij}r_i^tc_j^t)_{ij})$. Summing over all terminal transcripts gives exactly $S\ln2\,H_\nu(XY\mid\Pi(XY))$.

For a finite tree, telescoping proves $S\ln2\,H_\nu(XY\mid\Pi(XY))\le U(\one,\one)$. For an almost-surely terminating tree, truncate at increasing depths, discard the nonnegative unfinished potentials, and pass to the limit over completed leaves. Finally subtract conditional entropy from $H(\nu)=\HH(W)/(S\ln2)$.
\end{proof}

This is the sufficient, full-signal direction of the external concealed-information framework \cite{BGPW2015,FHLY2019}. Its direct proof avoids requiring an optimal continuation function or proving that a restricted family of weak signals is complete.

\subsection{The rational polynomial}
Use the weight matrix
\begin{equation}
\label{eq:W}
W=\Wmat.
\end{equation}
This is \emph{not} the function matrix: $W_{22}=0$, whereas $F(2,2)=1$.
For $\alpha=(\alpha_1,\alpha_2,\alpha_3)\in\mathbb N^3$, $|\alpha|=8$, write
\[
B_\alpha^{(8)}(r)=\frac{8!}{\alpha_1!\alpha_2!\alpha_3!}
 r_1^{\alpha_1}r_2^{\alpha_2}r_3^{\alpha_3},
\qquad r\in\Delta_2.
\]
There are $45$ such indices. The lower-bound certificate supplies the corresponding $45\times45$ rational coefficient array.\zenodopath{lower/support/baseline/potential.json} It defines
\begin{equation}
\label{eq:poly}
P_8(r,c)=\sum_{|\alpha|=|\beta|=8}
 q_{\alpha\beta}B_\alpha^{(8)}(r)B_\beta^{(8)}(c).
\end{equation}
It is separately of degree eight. The index order and a checksum identifying the exact file are recorded in Appendix~\ref{app:repro}.

\begin{proposition}[Finite polynomial certificate]
\label{prop:poly}
The rational polynomial \eqref{eq:poly} has the following globally verified properties on $\Delta_2\times\Delta_2$:
\begin{equation}
\label{eq:polychecks}
D_r^2P_8\preceq0,
\qquad D_c^2P_8\preceq0,
\qquad
P_8\ge E_W\quad\text{on every monochromatic terminal rectangle}.
\end{equation}
The Hessians are restricted to the respective simplex tangent coordinates.
\end{proposition}
\begin{proof}[Computer-assisted proof]
The curvature test restricts the matrix polynomial $-D_r^2P_8$ to a finite covering by products of triangles. Each restriction is expressed in a Bernstein basis. The verifier checks that every $2\times2$ matrix coefficient is positive semidefinite, using exact integer signs of both diagonal entries and the determinant. Bernstein basis weights are nonnegative, so the complete matrix polynomial is positive semidefinite throughout every accepted cell. Subdivision identities are exact; the accepted cells cover the product domain. The check visits $68{,}267$ cells and accepts $34{,}134$ leaves, with maximum subdivision depth $29$. Transposition symmetry gives the column inequality.

For the terminal bound, it is enough to check $R_+,R_-$, the middle row and column, and the two singleton zero rectangles. On a cell of a nontrivial terminal domain, a positive rational reference distribution $q$ gives the entropy tangent bound
\begin{equation}
\label{eq:entropy-tangent}
\HH(w)\le\sum_iw_i\ln(1/q_i).
\end{equation}
The difference between $P_8$ and the right side is a polynomial. Its Bernstein coefficients are checked with rational subdivision and outward logarithm intervals. Symmetry reduces the nontrivial cases to one overlapping rectangle and the middle row. These checks visit $5{,}723$ and $1{,}587$ cells, respectively, accepting $2{,}862$ and $794$ leaves. The singleton checks are exact. Boundary points are covered by the same closed cells.

The independent verifier and data are part of the theorem's certificate. It invokes no optimizer. The arithmetic assumptions and reproduction procedure appear in Appendix~\ref{app:repro}. Floating point may guide the subdivision order, but no sampled sign is used to accept a cell.
\end{proof}

\subsection{Three compatible rational--logarithmic corrections}
\label{sec:corrections}
The polynomial is a certified background, not a proposed exact formula for the
optimal concealed entropy. We improve it by three explicitly specified,
overlapping interior corrections. Put
\[
a(r)=r_1-r_3,\quad b(r)=r_1+r_3,\quad b_0=\frac{48}{73},\qquad
\psi(s)=\ln(1+s)-\frac{s}{1+s}\quad(0\le s\le1).
\]
Then
\[
\psi(0)=\psi'(0)=0,\qquad
\psi'(s)=\frac{s}{(1+s)^2},\qquad
\psi''(s)=\frac{1-s}{(1+s)^3}.
\]
For $[x]_+=\max\{x,0\}$ define
\begin{align}
 s_0(r)&=\left[1-\left(\frac{a(r)}{3/5}\right)^2
                  -\left(\frac{b(r)-b_0}{1/4}\right)^2\right]_+,\notag\\
 s_\pm(r)&=\left[1-\left(\frac{a(r)\mp1/4}{3/10}\right)^2
                  -\left(\frac{b(r)-b_0}{1/5}\right)^2\right]_+,
 \qquad f_j(r)=\psi(s_j(r)).\label{eq:patch-regions}
\end{align}
The corrected supporting function is
\begin{equation}
\label{eq:P3}
\begin{split}
P_3(r,c)=P_8(r,c)
 &-\frac{11}{12500}f_0(r)f_0(c)\\
 &-\frac{9}{500000}\bigl(f_+(r)f_+(c)+f_-(r)f_-(c)\bigr).
\end{split}
\end{equation}
It is invariant under transposition and simultaneous endpoint exchange.
All subtracted terms are nonnegative. Lowering a valid concealed-entropy
potential increases the information lower bound, but concavity of the
\emph{combined} expression must be proved.

\begin{proposition}[Joint rational--logarithmic certificate]
\label{prop:patches}
The function $P_3$ is separately concave on $\Delta_2\times\Delta_2$ and
majorizes $E_W$ on every monochromatic terminal rectangle.
\end{proposition}
\begin{proof}[Analytical reduction and computer-assisted verification]
An ellipse with center $(a_j,b_0)$ and semiaxes $A_j,B_j$ is strictly inside
the likelihood simplex if
\[
|a_j|+\sqrt{A_j^2+B_j^2}<b_0,\qquad b_0+B_j<1.
\]
For the central ellipse the first left side is $13/20<48/73$;
for either side ellipse it is $1/4+\sqrt{13}/10<48/73$.
The second condition also holds. Thus every correction vanishes when
either likelihood vector is on the simplex boundary. At a terminal
rectangle at least one vector is on that boundary, since the full matrix
is nonconstant. The terminal inequalities are therefore unchanged.
This argument refers to the actual input rectangle, not just positive-prior inputs.

The identities $\psi(0)=\psi'(0)=0$ give matching first derivatives across
each elliptical support boundary. On a smooth piece, use the affine row
coordinates $(a,b)$, and set
\[
D_j=\diag(A_j^{-2},B_j^{-2}),\quad
z_j=(a-a_j,b-b_0)^{\mathsf T},\quad
\varepsilon_0=\frac{11}{12500},\quad
\varepsilon_+=\varepsilon_-=\frac9{500000}.
\]
There
\begin{equation}
\label{eq:patch-hessian}
D^2 f_j=4\psi''(s_j)(D_jz_j)(D_jz_j)^{\mathsf T}
            -2\psi'(s_j)D_j.
\end{equation}
The verifier proves the full joint matrix inequality
\begin{equation}
\label{eq:joint-patch}
-D_r^2P_8(r,c)+\sum_{j\in\{0,+,-\}}
\varepsilon_jf_j(c)D_r^2f_j(r)\succeq0.
\end{equation}
In particular, the favorable rank-one terms of \eqref{eq:patch-hessian}
are retained; independent curvature allowances for the three changes
would be unnecessarily restrictive.

The check uses exact integer Bernstein restrictions of the polynomial
Hessian, dyadic subdivision, and rational enclosures for all correction
entries, including cells that meet several supports. For $1\le x\le2$,
with $z=(x-1)/(x+1)$, its logarithm bounds are
\begin{equation}
\label{eq:log-series}
2\sum_{k=0}^{N-1}\frac{z^{2k+1}}{2k+1}
\le\ln x\le
2\sum_{k=0}^{N-1}\frac{z^{2k+1}}{2k+1}
+\frac{2z^{2N+1}}{(2N+1)(1-z^2)},\qquad N=18.
\end{equation}
For each polynomial Bernstein matrix coefficient, add the correction
interval. If the diagonal lower bounds are $A,C$ and the off-diagonal
interval is $[B_-,B_+]$, the exact acceptance conditions are
\[
A\ge0,\qquad C\ge0,\qquad
AC\ge\max\{B_-^2,B_+^2\}.
\]
These certify positive semidefiniteness throughout the cell. The completed
check visits $21{,}191$ cells, accepting $8{,}798$ active and $1{,}798$
inactive leaves, with maximum subdivision depth $35$. Transposition gives
column concavity. Matching first derivatives across the finitely many
support interfaces along a line completes the global concavity proof.
The verifier invokes no optimizer; Appendix~\ref{app:repro} specifies the deposited data and arithmetic assumptions.
\end{proof}

The side corrections genuinely compensate for the central change. With
only the central term of \eqref{eq:P3}, concavity fails at the rational state
\[
(a(r),b(r),a(c),b(c))=
\left(\frac{477}{100000},\frac{65563}{100000},
-\frac{497}{2500},\frac{32891}{50000}\right).
\]
For $d=(1,-601/4000)^{\mathsf T}$, outward evaluation gives
\begin{align*}
d^{\mathsf T}\{-D_r^2(P_8-\varepsilon_0f_0f_0)\}d
 &\in(-8.6044430,-8.6044429)\,10^{-6},\\
d^{\mathsf T}(-D_r^2P_3)d
 &\in(1.5260504,1.5260505)\,10^{-5}.
\end{align*}
The first sign disproves the central-only candidate; the second is a
local illustration, not a replacement for Proposition~\ref{prop:patches}.
The regions are curvature-selected correction regions. They are not
asserted to be exact contact regions of an optimal protocol.

Extend either $P_j$, $j\in\{8,3\}$, separately homogeneously:
\[
\widehat P_j(r,c)=s_rs_cP_j(r/s_r,c/s_c),\qquad
s_r=\sum_i r_i,\quad s_c=\sum_jc_j,
\]
with value zero if either sum is zero. The perspective extension preserves
separate concavity. At pairs of simplex vertices the terminal entropy is
zero, so separate concavity implies $P_j\ge0$ throughout. Consequently
\begin{equation}
\label{eq:potential-U}
U_3(r,c)=\min\{E_W(r,c),\widehat P_3(r,c)\}
\end{equation}
satisfies Lemma~\ref{lem:potential}. The minimum of concave functions is
concave, since its hypograph is the intersection of their convex hypographs.

\subsection{A rational hard prior and the exact endpoint}
The lower bound is slightly stronger after a positive diagonal change of likelihood coordinates. Put $\rho=24/25$ and
\[
W_\rho=\begin{pmatrix}0&1&\rho\\1&0&1\\\rho&1&0\end{pmatrix},
\qquad
D_\rho=\diag(\sqrt\rho,1/\sqrt\rho,\sqrt\rho).
\]
Then $W_\rho=D_\rho W D_\rho$. Composing \eqref{eq:potential-U} with $(r,c)\mapsto(D_\rho r,D_\rho c)$ gives a valid certificate for $W_\rho$. Positivity of the diagonal map preserves the actual support rectangles. The normalized hard prior is
\begin{equation}
\label{eq:hardprior}
\nu_* =\frac1{148}\begin{pmatrix}0&25&24\\25&0&25\\24&25&0\end{pmatrix}.
\end{equation}
The normalized initial likelihoods in the $W$ coordinates are
$v=(24,25,24)/73$. Direct rational evaluation gives
\begin{equation}
\label{eq:J-evaluation}
\frac{5329}{3552}P_8(v,v)
=\frac{\Jnum}{\Jden}=J_8.
\end{equation}
The prefactor is $\bigl(2\sqrt\rho+1/\sqrt\rho\bigr)^2/(4+2\rho)$, converting the homogeneous potential to a concealed-entropy bound per unit prior mass.

At $v$ the three arguments satisfy $s_0(v)=1$ and
$s_\pm(v)=11/36$. The corrected concealed entropy per unit prior mass is
therefore the exact number
\begin{equation}
\label{eq:J-corrected}
\begin{split}
J_* = J_8-\frac{5329}{3552}\Bigg[
 &\frac{11}{12500}\left(\ln2-\frac12\right)^2\\
 &+\frac9{250000}
   \left(\ln\frac{47}{36}-\frac{11}{47}\right)^2\Bigg].
\end{split}
\end{equation}
The entropy of \eqref{eq:hardprior} in nats is
\[
\ln(148/25)+\frac{12}{37}\ln(25/24).
\]
Consequently, Lemma~\ref{lem:potential} proves the following.
\begin{theorem}[Explicit hard-prior lower bound]
\label{thm:lower}
Every fully correct randomized protocol for $F$ satisfies
\begin{equation}
\label{eq:k-lower}
\ICcost{\nu_*}(\Pi)\ge k_*:=
\frac{\ln(148/25)+(12/37)\ln(25/24)-J_*}{\ln2}
=1.2331249058620917801\ldots.
\end{equation}
Hence $\ICrnd(F)\ge\Lstar$ as defined in \eqref{eq:maininterval}.
\end{theorem}
\begin{proof}
Equation~\eqref{eq:J-corrected} bounds the initial concealed potential per unit mass, and \eqref{eq:k-lower} follows by subtracting it from the prior entropy. Taking the infimum over the fully correct private-coin protocols gives $\ICrndmu{\nu_*}(F)\ge k_*$ and hence $\kappa\ge k_*$. Apply the distribution-first identity in Proposition~\ref{prop:threeclass}.
\end{proof}

A full-input hard prior can also be written explicitly. With $T=2^{k_*}$, set
\begin{equation}
\label{eq:fullhard}
\mu_* =\frac{\delta_{11}+\delta_{33}+T\nu_*}{2+T}.
\end{equation}
Then \eqref{eq:threeclass-mi} and \eqref{eq:k-lower} give $\ICcost{\mu_*}(\Pi)\ge\Lstar$ for every fully correct protocol. Neither $\nu_*$ nor $\mu_*$ is claimed to maximize the unrestricted distributional complexity.

The displayed digits evaluate the explicit certificate expression; they do not mean that the unknown value of $\ICrnd(F)$ is known to that precision.

\section{A finite protocol and its uniform upper certificate}
\label{sec:upper}
The upper certificate specifies an ordinary finite private-coin randomized protocol, not a physical-clock protocol and not an unverified limit. Its large decision graph is deposited with the verification code.\zenodopath{upper/data/pruned-tail.npz} This section describes how to interpret that data as a protocol and how the verified channel bounds imply the theorem.

\subsection{Interpreting the rational protocol data}
There are two equivalent descriptions of a local signal. In likelihood coordinates, an Alice signal can be specified by an exact convex decomposition
\begin{equation}
\label{eq:signal-decomposition}
r=\sum_b\lambda_b r^{(b)},\qquad
\lambda_b\ge0,\quad\sum_b\lambda_b=1.
\end{equation}
For $r_i>0$, Alice implements it by
\begin{equation}
\label{eq:signal-implementation}
\Pr[B=b\mid X=i,r]=\lambda_b\frac{r_i^{(b)}}{r_i}.
\end{equation}
These probabilities sum to one. Coordinates with $r_i=0$ cannot occur at
that state and may be assigned arbitrary rules. Bob's messages are
analogous. Alternatively, a local rule can be specified directly by its
rational conditional probabilities; the certificate uses both
representations.

A stored continuation is a complete conditional protocol, not merely an
entropy value at a design prior. It can follow any message for which it
remains correct on every reachable input. The stored continuations are
checked on the entire input rectangle; the design probabilities used to
choose them do not restrict the prior under which they may be run.

The protocol consists of a specified sixteen-stage prefix followed by
finite continuation protocols on a rational likelihood grid, with a
deterministic correct finish. A scheduled signal may be a no-op; it is not
necessarily a transmitted bit or an alternating communication round.
At $421$ cut states the certificate specifies a rational continuation-selection
rule. The authoritative rules are included in the upper-bound module.\zenodopath{upper/all-final-records.json} At all other cut
states, the continuation is prescribed by the stored decision graph.
For the certified protocol, every conditional probability is fixed at
its supplied rational value. The conditional tail directed acyclic graph
has $3{,}278{,}123$ nodes and $16{,}214$ aggregated terminal likelihood
states. These are data sizes, not a round count. Each finite message is
encoded by an ordinary prefix-free binary code, preserving external
information.

For verification, paths may be aggregated only when their input likelihood vectors are proportional. If $\Pr[t\mid z]=a_tg(z)$ within a group, the conditional distribution of $t$ given the group is independent of $z$. The group is therefore a sufficient statistic for the input, and aggregation leaves external information unchanged. Merely sharing an output label would not justify aggregation. The transmitted symmetry label and distinct continuation choices remain part of the transcript whenever their likelihood ratios differ.

\subsection{The reduced information radius}
The protocol uses the private-seed symmetrization of
Section~\ref{sec:reduction}: Alice samples the group element with her
private coins and sends its two-bit encoding. The verified transcript
channel includes this label, so the information calculation applies
directly to the private-coin implementation. Define a rational reference prior on $S_1$ by
\begin{equation}
\label{eq:upperprior}
x=\frac{287514831671}{10^{12}},
\qquad
\widehat\nu=
\begin{pmatrix}
0&(1-x)/4&x/2\\
(1-x)/4&0&(1-x)/4\\
x/2&(1-x)/4&0
\end{pmatrix}.
\end{equation}
Let $W_z$ be its transcript law on input $z$ and $Q=\sum_z\widehat\nu_zW_z$. The matrix $\widehat\nu$ is a reference distribution, not an asserted optimal prior. In particular, its zero center mass is not a promise.

\begin{proposition}[Finite upper certificate]
\label{prop:upper-certificate}
The supplied rational protocol, with the communicated seed just described, is fully correct on all nine inputs and uses only private coins. For its symmetrized reduced channel, the divergences from $Q$ have the strict upper bounds in Table~\ref{tab:radius}.
\end{proposition}
\begin{table}[htbp]
\centering
\caption{Input-wise divergence bounds for the \emph{reduced} channel, in bits. The center is checked even though it has zero reference mass.}
\label{tab:radius}
\begin{tabular}{@{}llr@{}}
\toprule
Input class & Inputs & Strict upper bound\\
\midrule
Adjacent one-inputs & $12,21,23,32$ & $1.251085696967958$\\
Cross one-inputs & $13,31$ & $1.251085696967991$\\
Center & $22$ & $0.216212811303620$\\
\bottomrule
\end{tabular}
\end{table}
\begin{proof}[Computer-assisted proof]
The independent verifier reconstructs the finite protocol from its rational local rules. It checks nonnegativity and normalization, the speaking party's locality, the exact likelihood identities, and monochromatic terminal supports. In particular, the two zero inputs and the center are checked structurally; the information calculation does not substitute support-only correctness.

It then propagates the conditional transcript laws with outward intervals and evaluates the three symmetry classes of $D(W_z\Vert Q)$. The base finite tail graph has approximately $3.28$ million nodes; exact path-factorization permits its compressed evaluation. The logarithm routine uses range reduction and an atanh series with an explicit remainder. Correctly rounded binary64 operations, outward endpoint rounding, and compilation without fast-math are the stated arithmetic assumptions. Large finite input data, exact checks, and source code are specified in Appendix~\ref{app:repro}. No numerical claim that $\widehat\nu$ is a capacity maximizer is used.
\end{proof}

\begin{theorem}[Uniform upper bound]
\label{thm:upper}
The external information complexity satisfies $\ICrnd(F)<2.130998318$.
\end{theorem}
\begin{proof}
For every prior $\nu$ supported on all seven one-inputs, including arbitrary mass at the center, \eqref{eq:radius} and Table~\ref{tab:radius} give
\[
\ICcost{\nu}(\Pi)<1.251085696967991.
\]
The non-strict bound with this rational decimal is already enough: by Proposition~\ref{prop:threeclass},
\[
\max_\mu \ICcost{\mu}(\Pi)
\le\log_2\!\left(2+2^{1.251085696967991}\right)
<2.130998318.
\]
For each prior, the infimum over private-coin protocols is at most this particular protocol's cost; maximizing over priors therefore gives the asserted bound on $\ICrnd(F)$. The verifier's unrounded transformed upper enclosure is $2.130998317219$. The displayed reduced per-class values are not full-channel divergences for the zero-input classes; the exact three-class transformation is what yields the full bound.
\end{proof}

Combining Theorems~\ref{thm:lower} and \ref{thm:upper} proves Theorem~\ref{thm:main}. The certified interval has width less than $0.009732151294$ bits.

\subsection{What the finite certificate does not prove}
The protocol's successful verification proves a uniform upper bound and all-input correctness. It does not prove that this finite protocol, its tail library, or its reference prior is optimal. The lower potential in Section~\ref{sec:lower} is also not claimed to be the exact continuation. The two independently justified bounds remain separated.

A numerical search may select probabilities or candidate continuations; the final bound must still be checked after rationalization. Increasing the number of rounds within an unchanged library, observing a small KKT residual, or stabilizing a grid calculation does not by itself close the global optimization problem. Appendix~\ref{app:additional} explains how local choices enter a global information-radius bound.

\section{Continuation pricing and supporting-plane contacts}
\label{sec:pricing}
\label{app:pricing}
This section gives exact local optimization identities for finite continuation libraries. These identities do not replace the global potential proof. Entropies and divergences in this section are in \emph{nats}.

For $d\ge1$, write $\Delta_{d-1}=\{a\in\mathbb R_{\ge0}^d:
\one^{\mathsf T}a=1\}$ for the probability simplex.

\subsection{The finite-library primal and dual}
Fix a prior $\nu$ at a protocol state and a finite collection of fully correct continuation channels $W_j(t\mid x,y)$, $1\le j\le J$. Suppose Bob selects $j$ with conditional probability $p_j(y)$ and then the parties run $W_j$. The choice $j$ is retained in the transcript. Initially assume each continuation is permitted on the entire remaining rectangle; support-restricted variants require the corresponding additional zero constraints.

For a nonnegative vector $a\in\R^{\mathcal Y}$ define
\begin{equation}
\label{eq:Psi}
\Psi_j(a)=\sum_t\HH\left(
  (\nu_{xy}a_yW_j(t\mid x,y))_{xy}\right).
\end{equation}
This function is continuous, concave, and positively homogeneous. The maximum concealed entropy obtainable by a first Bob message followed by this library is
\begin{equation}
\label{eq:master}
\Phi_{\mathcal W}(\nu)=
\max_{p_j\ge0,\ \sum_jp_j=\one}\ \sum_j\Psi_j(p_j).
\end{equation}
This is a convex optimization problem in the standard convention of maximizing a concave objective over a convex feasible set.

\begin{theorem}[Supporting-plane dual]
\label{thm:pricing-dual}
For the finite-library problem,
\begin{equation}
\label{eq:dual}
\Phi_{\mathcal W}(\nu)=
\min_{\lambda\in\R^{\mathcal Y}}
\left\{\one^{\mathsf T}\lambda:
\Psi_j(a)\le\lambda^{\mathsf T}a
\quad\text{for every }j\text{ and }a\in\Delta_{|\mathcal Y|-1}\right\}.
\end{equation}
\end{theorem}
\begin{proof}
Homogeneity extends each simplex inequality to all $a\ge0$. Summing it over a feasible allocation gives $\sum_j\Psi_j(p_j)\le\one^{\mathsf T}\lambda$, proving weak duality. Conversely, attach multiplier $\lambda$ to $\sum_jp_j=\one$. The supremum of the Lagrangian over nonnegative $p_j$ is finite precisely when each $\Psi_j(a)-\lambda^{\mathsf T}a\le0$ on the nonnegative cone. The feasible choice $p_j=\one/J$ is strictly positive, and the objective is finite and continuous. The finite-dimensional concave program therefore has strong Lagrange duality under the relative-interior condition \cite{BoydVandenberghe2004}. This gives \eqref{eq:dual}.
\end{proof}

For a candidate continuation $W$ define its \emph{price}, or supporting-plane violation, by
\begin{equation}
\label{eq:price}
\beta_W(\lambda)=\max_{a\in\Delta_{|\mathcal Y|-1}}
\{\Psi_W(a)-\lambda^{\mathsf T}a\}.
\end{equation}
A positive price violates the tested plane. If all prices in a tested finite library are at most $\varepsilon$, then $\lambda+\varepsilon\one$ is feasible in \eqref{eq:dual}, giving the certified library bound
\[
\Phi_{\mathcal W}(\nu)\le\one^{\mathsf T}\lambda+|\mathcal Y|\varepsilon.
\]
This statement controls only tested continuations.

At an exact optimum, a positive price for a new correct continuation permits a strict local improvement. For a witness $a$, introduce it with conditional probability $\varepsilon a_y$ and change the old probabilities to $(1-\varepsilon a_y)p_j(y)$. At differentiable active allocations, the old objective loses $\varepsilon\lambda^{\mathsf T}a+o(\varepsilon)$, while the new branch gains $\varepsilon\Psi_W(a)$. More robustly, a computer certificate may bound the entire old library by a feasible dual plane and exhibit a new feasible allocation above that bound, as in the globally weighted dual construction of Section~\ref{sec:global-family}.

\begin{proposition}[Number of first-message outcomes]
\label{prop:threeoutcomes}
Some optimizer of \eqref{eq:master} uses at most $|\mathcal Y|$ nonzero first-message outcomes. For $F$, at most three suffice.
\end{proposition}
\begin{proof}
Take an optimizer and write every nonzero allocation as $p_j=t_ja_j$ with $a_j\in\Delta_{|\mathcal Y|-1}$ and $t_j>0$. Fix these directions and optimize the weights by the linear program
\[
\max_{s_j\ge0}\sum_js_j\Psi_j(a_j),
\qquad \sum_js_ja_j=\one.
\]
The original weights are feasible. Any other feasible weights still define a valid allocation in \eqref{eq:master}, so the optimum is unchanged. A basic optimal solution has at most $|\mathcal Y|$ positive weights.
\end{proof}
The coefficients $t_j$ sum to $|\mathcal Y|$, not one. They are resource weights; the executable conditional probabilities are $p_j(y)=t_ja_j(y)$.

\subsection{Binary entropy and a graph-Laplacian Hessian}
For one fixed continuation, put
\[
b_{ty}=\sum_x\nu_{xy}W(t\mid x,y),
\qquad
\ell_y=\sum_t\HH\bigl((\nu_{xy}W(t\mid x,y))_x\bigr).
\]
Entropy decomposition into the column label and the input within a column gives
\begin{equation}
\label{eq:Psi-decomp}
\Psi_W(a)=\sum_t\HH\bigl((b_{ty}a_y)_y\bigr)+\ell^{\mathsf T}a.
\end{equation}
Assume now that $\nu$ is positive on the six off-diagonal entries and zero on the diagonal. At a terminal transcript of positive $\nu$-probability, at most two column labels occur. A one-rectangle containing all three columns can contain only the middle row; its middle cell has zero prior mass. Every other one-rectangle uses at most two columns, and zero-output rectangles have zero prior mass. Thus every nonlinear entropy term in \eqref{eq:Psi-decomp} is binary.

Create a graph on the three columns, with an edge $ij$ for each transcript for which $b_{ti},b_{tj}>0$. Multiple transcripts may give the same edge. For positive $a$, let
\[
\omega_t(a)=\frac{b_{ti}a_i\,b_{tj}a_j}{b_{ti}a_i+b_{tj}a_j}.
\]
\begin{proposition}[Graph curvature and unique contacts]
\label{prop:graph}
For every direction $d$,
\begin{equation}
\label{eq:graph-hessian}
-d^{\mathsf T}D^2\Psi_W(a)d
=\sum_{t:\supp(b_t)=\{i,j\}}
\omega_t(a)\left(\frac{d_i}{a_i}-\frac{d_j}{a_j}\right)^2.
\end{equation}
If the support graph is connected, then for each finite $\lambda$ the pricing problem \eqref{eq:price} has a unique interior optimizer $a_W(\lambda)$. It satisfies
\begin{equation}
\label{eq:pricing-kkt}
\nabla\Psi_W(a_W)=\lambda+\beta_W(\lambda)\one,
\qquad \one^{\mathsf T}a_W=1.
\end{equation}
\end{proposition}
\begin{proof}
Differentiate each two-coordinate homogeneous entropy in \eqref{eq:Psi-decomp}. Its negative Hessian quadratic form is exactly the corresponding term of \eqref{eq:graph-hessian}; the $\ell$ term is affine. If the graph is connected, the kernel consists of $d_i/a_i$ constant across all columns, hence $d\in\operatorname{span}(a)$. On the simplex tangent space this kernel is trivial, proving strict concavity in the interior.

No maximizer can lie on the boundary. For the set of positive coordinates at a boundary point, connectedness gives an edge to an absent coordinate. Transferring a small mass $\varepsilon$ to that coordinate creates a positive term of order $\varepsilon\ln(1/\varepsilon)$ in its binary entropy. The other terms have at worst linear loss along this transfer; the finite plane term is also linear. Thus the objective increases. Existence follows from compactness and continuity; strict concavity then gives uniqueness. The interior KKT equations have a common simplex multiplier. Dotting the gradient equation with $a_W$ and using Euler's homogeneous identity shows that this multiplier is $\beta_W(\lambda)$.
\end{proof}
If the graph is disconnected, the same formula describes independent connected components. The connected uniqueness conclusion should not be applied without this hypothesis.

\subsection{Global tangent inequalities and flat pieces}
For positive $\bar a$, define
\[
s_t(a)=\sum_yb_{ty}a_y,
\qquad q_t^a(y)=b_{ty}a_y/s_t(a).
\]
Terms with $s_t(a)=0$ contribute zero. Expanding homogeneous entropy in \eqref{eq:Psi-decomp} yields the exact tangent-gap identity
\begin{equation}
\label{eq:tangentgap}
\nabla\Psi_W(\bar a)^{\mathsf T}a-\Psi_W(a)
=\sum_t s_t(a)\KLnat(q_t^a\Vert q_t^{\bar a})\ge0.
\end{equation}
Thus a tangent plane majorizes $\Psi_W$ on the entire cone, not merely near $\bar a$. Under the connectedness hypothesis, equality on the simplex requires $a=\bar a$.

For connected channels, the pricing residual is differentiable in the plane
with $\nabla_\lambda\beta_j(\lambda)=-a_j(\lambda)$.
The local contact system can therefore be written
\begin{equation}
\label{eq:contact-system}
\begin{gathered}
\beta_j(\lambda)\le0,\quad t_j\ge0,\quad t_j\beta_j(\lambda)=0,
\qquad \sum_jt_ja_j(\lambda)=\one,\\
\nabla\Psi_j(a_j)=\lambda+\beta_j(\lambda)\one.
\end{gathered}
\end{equation}
Every used continuation has price zero and shares the same supporting plane. Only three need be active for the present function.

To express the resulting flat contact geometry, allow a general resource vector $b\ge0$ and define
\[
G_{\mathcal W}(b)=\max_{p_j\ge0,\ \sum_jp_j=b}\sum_j\Psi_j(p_j).
\]
If a dual-feasible plane touches channels at directions $a_j$, then
\begin{equation}
\label{eq:flatcone}
G_{\mathcal W}\!\left(\sum_jt_ja_j\right)
=\lambda^{\mathsf T}\sum_jt_ja_j\qquad(t_j\ge0).
\end{equation}
The upper inequality is dual feasibility; allocating $p_j=t_ja_j$ attains the reverse inequality. One fixed connected continuation has a unique simplex contact; nontrivial flat pieces arise by combining contacts of different continuations.

For a feasible contact allocation satisfying $\sum_jt_ja_j=\one$ and a
normalized column likelihood $c$, the child likelihoods and decomposition
weights are
\[
c^{(j)}=\frac{c\odot a_j}{c^{\mathsf T}a_j},
\qquad \theta_j=t_jc^{\mathsf T}a_j,
\qquad c=\sum_j\theta_jc^{(j)}.
\]
Terms with $c^{\mathsf T}a_j=0$ are omitted. These weights sum to one but are
not generally the signal probabilities under the prior. The equations describe exact contacts for the finite continuation envelope. They do not establish the other party's concavity for an arbitrary globally proposed envelope.

\subsection{What a new contact does and does not certify}
A positive price for a new continuation identifies a missing local option.
It does not by itself improve the worst-prior cost of the complete protocol:
the benefit must be weighted by the probability of reaching this state,
and the change can shift the global maximizing prior. A numerical local
improvement is therefore used only to propose an additional fully correct
continuation. Appendix~\ref{app:additional} formulates the joint optimization of its
selection with the other menus at a fixed cut and the corresponding all-input
information-radius bound.

Nor does absence of detected positive prices certify the unrestricted
problem. It certifies only the specified library when its entire dual
inequalities have been verified. The unrestricted problem requires
compatibility under both parties' signal splittings, including continuations
not yet present in the library.

\section{Global continuation optimization and feedback compatibility}
\label{sec:global-family}
\label{app:additional}
Local continuation improvements must be coordinated by their reach probabilities
and by a common worst-prior test. This appendix formulates optimization within a
finite continuation family, derives bounds from local supporting planes, and
gives general criteria for compatibility under both parties' signal splittings.

\subsection{An affine channel at a fixed transcript cut}
Fix a finite prefix of a fully correct protocol. At each unfinished transcript
$h$, write $\alpha_h(z)=\Pr[h\mid Z=z]$, where $Z=(X,Y)$. Specify the next
speaker $\sigma_h$ and a finite library of fully correct conditional
continuations $\Pi_{hj}$, with channels $W_{hj}(t\mid z)$. The speaker chooses
$j$ with probability $p_{hj}(z_{\sigma_h})$, subject to
\[
p_{hj}(i)\ge0,\qquad \sum_jp_{hj}(i)=1.
\]
The complete channel on that branch is
\begin{equation}
\label{eq:global-affine}
W_p(h,j,t\mid z)=\alpha_h(z)p_{hj}(z_{\sigma_h})W_{hj}(t\mid z).
\end{equation}
Completed prefix transcripts remain unchanged. The channel is affine in all
new selection probabilities: the variable messages occur at disjoint
transcripts, each followed by a fixed continuation. Optimizing several
successive variable messages on the same branch would generally introduce
products of variables and would not be the convex problem below.

\begin{theorem}[Exact optimization within a finite family]
\label{thm:finite-global}
For the specified family $\mathcal L$, let $S_1=F^{-1}(1)$. Its optimal
uniform reduced cost satisfies
\begin{align}
k_{\mathcal L}
&:=\min_p\max_{\nu\in\Delta(S_1)}\ICcost{\nu}(\Pi_p)\notag\\
&=\min_{p,Q}\max_{z\in S_1}D(W_p(\cdot\mid z)\Vert Q)
\label{eq:global-radius}\\
&=\max_{\nu\in\Delta(S_1)}\min_p\ICcost{\nu}(\Pi_p).
\label{eq:finite-minimax}
\end{align}
The epigraph formulation of \eqref{eq:global-radius} is convex.
\end{theorem}
\begin{proof}
For each fixed channel, capacity equals its information radius: the
maximizing-prior conditions \eqref{eq:capacity-kkt} provide a reference law
achieving equality, while \eqref{eq:radius} supplies the opposite inequality.
Relative entropy is jointly convex and \eqref{eq:global-affine} is affine.
For the minimax statement, mutual information is convex in a channel and
concave in its prior. Both parameter domains are compact convex sets, and
these finite-channel costs are continuous. The convex--concave minimax
theorem therefore applies \cite{Sion1958}; the standard convexity facts are
also discussed in \cite{CoverThomas2006,BoydVandenberghe2004}.
\end{proof}
The theorem is a property of $\mathcal L$, not a redefinition of
$\ICrnd(F)$. The center belongs to the outer simplex throughout.
For a fixed test prior set
\[
a_h=\sum_z\nu(z)\alpha_h(z),\qquad
\nu_h(z)=\frac{\nu(z)\alpha_h(z)}{a_h}\quad(a_h>0).
\]
The inner optimization decomposes as
\begin{equation}
\label{eq:global-decomposition}
\min_p\ICcost{\nu}(\Pi_p)
=I_\nu(Z:H_{\rm prefix})+
\sum_h a_h\min_{p_h}I_{\nu_h}(Z:\Pi_{h,p_h}(Z)).
\end{equation}
Zero-probability branches still have correct continuations but contribute
zero to this objective. Thus posterior problems remain useful; they are
coordinated by one outer prior and weighted by their actual reach probabilities.
For a symmetry-closed family a maximizing prior may be taken as
\[
\nu=\begin{pmatrix}0&b&c\\b&d&b\\c&b&0\end{pmatrix},
\qquad 4b+2c+d=1.
\]
There are two free outer parameters; $d=0$ is not imposed by the theory.

\subsection{A bound on improvement within a finite family}
At each variable cut state, the local problem in
Appendix~\ref{sec:pricing} has concealed-entropy optimum $\Phi_h$ in nats.
At positive rational tangent directions $a_{hj}$,
\[
\Psi_{hj}(a)\le\nabla\Psi_{hj}(a_{hj})^{\mathsf T}a
\qquad(a\ge0).
\]
A rational vector $\lambda_h$ dominating all these gradients coordinatewise
is a common global majorant, and gives
$\Phi_h\le\one^{\mathsf T}\lambda_h$. Fix a feasible protocol
$\Pi_0\in\mathcal L$, and let $g_h$ be the concealed entropy of its
allocation at that state under the test prior. Then
\begin{equation}
\label{eq:family-error}
\varepsilon_{\mathcal L}
=\frac1{\ln2}\sum_h a_h
\bigl(\one^{\mathsf T}\lambda_h-g_h\bigr)
\end{equation}
upper-bounds the possible reduction of information cost at this test prior
by optimizing all those menus. Transcripts with identical likelihood ratios
and the same menu may be aggregated by summing their reach weights.

Let $W_z^0$ be the reduced transcript law of $\Pi_0$ on input $z$,
and let $Q$ be a reference law dominating these laws. For every fixed test
prior $\nu\in\Delta(S_1)$, the preceding bound gives
\[
\ICcost{\nu}(\Pi_0)-\varepsilon_{\mathcal L}
\le k_{\mathcal L}
\le\max_{z\in S_1}D(W_z^0\Vert Q).
\]
The left inequality tests the common prior $\nu$ against every member of
$\mathcal L$; the right inequality uses the feasible protocol $\Pi_0$ and
\eqref{eq:radius}. A lower bound on $k_{\mathcal L}$ applies to the specified
family: restricting the permitted continuations can increase the optimum.

\subsection{The globally coordinated feedback law}
Choose continuation reference laws $Q_{hj}$ and put
\[
d_{hj}(z)=D(W_{hj}(\cdot\mid z)\Vert Q_{hj}).
\]
With reference message probabilities $q_{hj}$, the local radius contribution is
\[
\sum_jp_{hj}(i)
\left[\log_2\frac{p_{hj}(i)}{q_{hj}}+d_{hj}(z)\right],
\qquad i=z_{\sigma_h}.
\]
Let $\theta_z$ be the multipliers of the \emph{global} input-wise
inequalities. At a state with $\sum_z\theta_z\alpha_h(z)>0$, they induce
\[
\omega_h(z)=\frac{\theta_z\alpha_h(z)}{\sum_u\theta_u\alpha_h(u)},
\qquad
\overline d_{hj}(i)=\E_{\omega_h}[d_{hj}(Z)\mid Z_{\sigma_h}=i].
\]
The conditional expectations and the first stationarity equation below
apply only to inputs $i$ with $\omega_h(Z_{\sigma_h}=i)>0$, on active
alternatives with finite continuation costs. Inputs of zero $\omega_h$-mass
do not constrain these stationarity equations; their rules must still
specify correct continuations.
Minimizing the weighted radius contribution over a positive probability
vector gives the exact stationarity equations
\begin{equation}
\label{eq:feedback-law}
p_{hj}(i)=
\frac{q_{hj}2^{-\overline d_{hj}(i)}}
{\sum_k q_{hk}2^{-\overline d_{hk}(i)}},
\qquad
q_{hj}=\sum_i\omega_h(Z_{\sigma_h}=i)p_{hj}(i),
\end{equation}
where the second equality assumes that the reference message weights are
optimized as well. The first follows by a Lagrange multiplier for
$\sum_jp_{hj}(i)=1$; the second is the corresponding mixture consistency
condition. Zeros require one-sided optimality conditions.

The exponential rule favors continuations with small future leakage while
paying for information revealed by choosing them. It is a short
state-dependent selection law, but not a closed-form optimal protocol:
the continuation costs, references, and global multipliers must also be
specified consistently. A finite library cannot be declared complete
merely because its own stationarity equations hold.

\subsection{Why smooth reoptimization can violate the other player's concavity}
Fix a finite Bob-first menu. Let $u$ be free coordinates for its probability
rules and let $\mathcal F(r,c,u)$ be its homogeneous concealed entropy.
Write $G_B(r,c)=\max_u\mathcal F(r,c,u)$. Reparameterizing by allocated column
vectors shows that $G_B$ is concave in $c$. Concavity in $r$ is not automatic.

\begin{proposition}[Opposite-player curvature]
\label{prop:schur-feedback}
Suppose the active constraints are fixed, the maximizing $u_*(r,c)$ is unique
and interior in its free coordinates, and $A=-\mathcal F_{uu}\succ0$.
On this regular region,
\begin{equation}
\label{eq:reoptimization-curvature}
D_r^2G_B=\mathcal F_{rr}
+\mathcal F_{ru}A^{-1}\mathcal F_{ur}.
\end{equation}
Row concavity is equivalent to
\begin{equation}
\label{eq:schur-feedback}
\begin{pmatrix}
-\mathcal F_{rr}&\mathcal F_{ru}\\
\mathcal F_{ur}&A
\end{pmatrix}\succeq0.
\end{equation}
\end{proposition}
\begin{proof}
Differentiate $\mathcal F_u(r,c,u_*)=0$ and use the envelope identity.
The Schur complement of $A$ gives the equivalence.
\end{proof}
For a fixed rule, $\mathcal F_{rr}\preceq0$. Reoptimization adds a
positive-semidefinite term. Smoothness of the selector thus does not
establish compatibility. At nonsmooth joins and rank changes, the full
signal inequalities replace the differential test.

The full signal condition also gives a test without differentiability
assumptions. For an Alice splitting $r=\sum_b r_b$ at fixed $c$, define
\[
\delta=\sum_b G_B(r_b,c)-G_B(r,c),\qquad
g=E(r,c)-\sum_b E(r_b,c)\ge0.
\]
These are homogeneous values on the unnormalized child likelihood vectors.
Separate concavity of $G_B+\eta E$ requires $\delta\le\eta g$.
A positive child excess over a valid parent upper bound therefore disproves
Alice concavity of $G_B$; when $g>0$, any compatible correction must satisfy
$\eta\ge\delta/g$.

If such a violation is established, it identifies a useful composite
continuation: an Alice signal followed by posterior-specific Bob menu
allocations. Once fixed, this composite is a correct continuation and can
be added to a subsequent globally coordinated optimization without
destroying the affine form \eqref{eq:global-affine}.

\subsection{Limiting feedback and a possible two-sided match}
At each replanning stage, the current speaker privately draws a fresh
input-independent Bernoulli$(\delta)$ stopping bit, with $0<\delta<1$,
and transmits it. On the stopping outcome the parties execute a fully
correct backup; otherwise they continue to replan. The stopping decision
remains in the transcript. This ensures almost-sure termination of the
feedback controller. A conditional reference law then gives the radius
recursion
\begin{equation}
\label{eq:feedback-radius}
\begin{split}
D_z(s)={}&\delta d_*^z(s)\\
&+(1-\delta)\left[D(p^z(s)\Vert q(s))+
\sum_b p_b^z(s)D_z(s_b)\right].
\end{split}
\end{equation}
Here $d_*^z$ is the backup divergence, $p^z$ the actual hint distribution,
and $q$ the reference hint distribution. The stopping bit uses the same law
in the actual and reference processes and so adds no divergence. A bounded
supersolution to \eqref{eq:feedback-radius}, with valid exit bounds, is a
uniform certificate; the geometric tail removes the remainder. With bounded
local costs, a uniform residual $\varepsilon$ entails an error at most
$\varepsilon/\delta$.

This safeguard is not an exactness argument. At a fixed prior, if the backup
cost is $a_\nu$ and the distributional optimum is
$\ICrndmu{\nu}(F)$, an independent, announced initial backup choice of probability
$\delta$ gives
\[
\ICcost{\nu}(\Pi_\delta)-\ICrndmu{\nu}(F)
\ge\delta\bigl(a_\nu-\ICrndmu{\nu}(F)\bigr).
\]
A suboptimal backup with fixed positive selection probability imposes a
nonzero bias. A limiting construction must remove that bias or provide a
different controlled completion; no unbounded controller is used in the
finite upper certificate of this paper.

A useful conditional test for a matching lower bound is the following.
Suppose an achievable, separately homogeneous concealed value $G$ satisfies
$0\le G\le E$ and $G=E$ at terminal rectangles. If $G+\eta E$ is separately
concave for some $\eta\ge0$, then
\[
U=\frac{G+\eta E}{1+\eta}
\]
is a valid potential. If the selected continuation costs $a_\nu$ at a
normalized prior $\nu$, Lemma~\ref{lem:potential} yields
\begin{equation}
\label{eq:two-sided-achievable}
\frac{a_\nu}{1+\eta}\le\ICrndmu{\nu}(F)\le a_\nu.
\end{equation}
For $\eta=0$ these bounds match. Applying this criterion requires an
achievable concealed value and a proof of the full signal inequalities for
both parties; smoothness or stationarity of a selector alone is insufficient.

\section{Conditional potential bounds at a transcript cut}
\label{sec:prefix}
A fixed opening determines a distribution of conditional priors at its
transcript cut. Entropy potentials at those priors bound every fully correct
continuation and hence the possible improvement from replacing the remaining
tails.

\begin{lemma}[Conditional potential bound at a transcript cut]
\label{lem:cut-bound}
Fix a finite opening and a prior $\nu$. Let $T$ be its cut transcript,
including branches that have already terminated. At each $h$, let $u_h$
be an upper bound in bits on the entropy that any fully correct continuation
can conceal under the conditional prior $\nu_h$. Then every extension
$\Pi$ of this opening satisfies
\begin{equation}
\label{eq:prefix-bound}
\ICcost{\nu}(\Pi)\ge H(\nu)-\sum_h\Pr_\nu[T=h]u_h.
\end{equation}
\end{lemma}
\begin{proof}
The complete transcript retains $T$. Its final conditional entropy is the
average over $h$ of the conditional entropy left by the continuation, which
is bounded by $u_h$. Subtract from $H(\nu)$.
\end{proof}
An entropy potential valid for a conditional prior provides the corresponding
bound $u_h$. At a fixed transcript, unused private randomness can be locally
resampled conditional on the party's own input and that transcript;
likelihood factorization therefore permits the usual continuation bound.
No restriction to a finite continuation library is imposed.

\subsection{Reach-weighted improvement budgets}
For a given protocol and a disjoint cut, let its current continuation conceal
$g_h$ bits at $h$, and let a valid potential bound the optimal concealed
entropy by $u_h$. At a fixed starting prior, replacing only these tails can
improve information cost by at most
\begin{equation}
\label{eq:improvement-budget}
\sum_h\Pr[h](u_h-g_h).
\end{equation}
The proof subtracts the current remaining entropy from the upper estimate
in Lemma~\ref{lem:cut-bound}. This bound can guide where to enlarge a
continuation family. A large budget need not be attainable: it can reflect
a loose potential. A small budget, however, rules out a large gain from
that set of changes. A subsequent uniform information-radius check is
still needed to convert design-prior improvements into a better upper
bound on $\ICrnd(F)$.

\section{A product-prior entropy certificate}
\label{app:product-prior}
This appendix records the one-copy certificate used in
Section~\ref{sec:direct-sums}. It is independent of the stronger
prior-free lower potential in Appendix~\ref{sec:lower}.
All entropies and divergences here are in bits.

\begin{proposition}[Product-prior lower bound]
\label{prop:product-prior}
Let $p=(9/20,1/10,9/20)$ and $\mu_*=p\otimes p$. Then
\[
 b_{\mu_*}(F)\ge
 \beta:=2H(p)-\frac{179}{250}
 =2h_2(1/10)+\frac{271}{250}
 =2.0219911871785624425\ldots>2.02.
\]
\end{proposition}
\begin{proof}
For the orientation $F=\Fmat$, define the weight matrix
\[
 W=\begin{pmatrix}
 0&7/5&4/5\\
 7/5&14&7/5\\
 4/5&7/5&0
 \end{pmatrix}.
\]
We first show that every product distribution $q=q_X\otimes q_Y$
supported on a monochromatic rectangle satisfies
\begin{equation}\label{eq:product-weight-entropy}
 H(q_X)+H(q_Y)\le\E_q W(X,Y).
\end{equation}
The maximal one-rectangles are the two overlapping $2\times2$ rectangles,
the middle row, and the middle column. On either $2\times2$ rectangle,
let $u$ and $v$ denote the probabilities of row and column label $2$.
Then \eqref{eq:product-weight-entropy} reduces to
\begin{equation}\label{eq:product-binary-envelope}
 h_2(u)+h_2(v)\le\frac45+\frac35(u+v)+12uv,
 \qquad 0\le u,v\le1.
\end{equation}
A finite exact certificate for this inequality is given below.
On the middle row, the entropy variational inequality gives
\[
 H(q_Y)\le\sum_yq_Y(y)W_{2y}
       +\log_2\left(\sum_y2^{-W_{2y}}\right)
       \le\sum_yq_Y(y)W_{2y},
\]
since $2\cdot2^{-7/5}+2^{-14}<1$. The middle column is identical.
The zero-rectangles are singletons with weight zero. This proves
\eqref{eq:product-weight-entropy} for all monochromatic rectangles,
including their subrectangles.

Let $T$ be the transcript of any fully correct private-coin protocol.
The product prior and the likelihood factorization
\eqref{eq:factorization} imply $X\perp Y\mid T$.
Each posterior is supported on a monochromatic rectangle, so
\eqref{eq:product-weight-entropy} gives
\[
 H(X,Y\mid T)
 =H(X\mid T)+H(Y\mid T)
 \le\E_{\mu_*}W(X,Y)=\frac{179}{250}.
\]
The final equality follows by summing the two corner weights, four adjacent
weights, and the center weight:
\[
 2\left(\frac9{20}\right)^2\frac45
 +4\left(\frac9{20}\right)\left(\frac1{10}\right)\frac75
 +\left(\frac1{10}\right)^2 14=\frac{179}{250}.
\]
Subtracting from $H(X,Y)=2H(p)$ and taking the protocol infimum proves
the bound. The strict numerical comparison follows from the exact
integer test stated below.
\end{proof}

\subsection{Finite certificate for the binary entropy inequality}
For $k=0,\ldots,6$, the binary entropy variational formula gives
\[
 h_2(u)\le ku+\log_2(1+2^{-k})\le ku+c_k,
\]
where the following values are exact rationals:
\[
 (c_0,\ldots,c_6)=
 \left(1,\frac{117}{200},\frac{161}{500},\frac{17}{100},
             \frac7{80},\frac{111}{2500},\frac{14}{625}\right).
\]
For $c_k=a/b$, its validity is checked using integers:
\[
 (2^k+1)^b\le2^{kb+a}.
\]
Set $g(u)=\min_{0\le k\le6}(ku+c_k)$.
It suffices to consider $u,v\le1/2$: reflection about $1/2$ preserves
binary entropy and decreases the right-hand side of
\eqref{eq:product-binary-envelope}.
On each rectangle between consecutive intersections of the affine pieces
of $g$, the expression
\[
 g(u)+g(v)-\frac35(u+v)-12uv
\]
is bilinear and has its maximum at a corner. Enumerating these finitely
many rational corners gives exactly
\[
 \max_{0\le u,v\le1/2}
 \left[g(u)+g(v)-\frac35(u+v)-12uv\right]
 =\frac{79783}{100000}<\frac45.
\]
This establishes \eqref{eq:product-binary-envelope}.
The middle-row estimate follows, without numerical logarithms, from
$2^7>(5/2)^5$ and $4/5+2^{-14}<1$.
Finally,
\[
 10^{250}>2^{117}9^{225}
 \quad\Longrightarrow\quad
 h_2(1/10)>\frac{117}{250}
 \quad\Longrightarrow\quad\beta>2.02.
\]
These finite tests use only the displayed integer and rational inequalities;
floating-point logarithms are needed only to display the final decimal value.

\section{Computational certificates and reproducibility}
\label{app:repro}

The authoritative computational package supporting the certified lower and
upper bounds in Theorem~\ref{thm:main}, and the exhaustive Boolean
$3\times3$ classification in Theorem~\ref{thm:classification}, is deposited
on Zenodo~\cite{WunderlichZenodo2026}. 

\subsection{Scope of the deposited checks}
The archive has three independent modules.\zenodopath{lower/, upper/, classification/}
\begin{description}[leftmargin=2.2em,style=nextline]
\item[Lower-bound module.]
It verifies the degree-eight rational polynomial background, the joint
separate-concavity inequalities for the three rational--logarithmic
corrections, and an outward interval evaluation of the endpoint
\(L_*\).

\item[Upper-bound module.]
It reconstructs the rational protocol prefix, checks the
$3{,}278{,}123$-node finite tail DAG, verifies speaker locality and
monochromatic terminal supports, applies the $421$ deployed continuation
rules, and obtains an outward enclosure of the transcript divergences
uniformly over input distributions.

\item[Classification module.]
It enumerates all $512$ Boolean $3\times3$ matrices, their $36$ row/column
orbits and $13$ symmetry families, and independently audits the
rectangle-separated witnesses and deterministic protocol trees used in the
classification.
\end{description}

The principal upper-bound data file is the finite tail DAG.\zenodopath{upper/data/pruned-tail.npz}
The coefficient order for the polynomial $P_8$ is
\[
(\alpha_1,\alpha_2,\alpha_3)=(i,j,8-i-j),\quad
 i=0,\ldots,8,\quad j=0,\ldots,8-i,
\]
with $i$ as the outer loop; the integer coefficient array is divided by
$10^{13}$. The archive-wide SHA-256 manifest identifies every deposited
file.\zenodopath{SHA256SUMS.txt}
Checksums identify the certificate inputs; they do not replace the
mathematical arguments connecting successful verification to the stated
bounds.

\subsection{Reproduction procedure}
The root README gives the installation and execution commands.\zenodopath{README.md}
Python~3 with NumPy, mpmath, and Numba is sufficient for the lower- and
upper-bound certificates; the classification module uses only the Python
standard library. A complete run of these three modules executes the
classification, lower-bound, and upper-bound verifiers and generates fresh
runtime outputs. No previously generated logs, summaries, or caches are
required: runtime caches are regenerated from the supplied authoritative
data and used during verification. These generated outputs are not included
in the deposit.

The lower verifier first checks the global polynomial certificate, then the
joint correction, and finally the interval evaluation of the lower endpoint.
The upper verifier reconstructs the prefix from rational source data, checks
the tail graph and all continuation rules, and then performs the information-
radius calculation. The classification verifier recomputes the tables and
witness audits from source rather than reading pre-generated classifications.

\subsection{Arithmetic assumptions}
The polynomial curvature tests use exact arbitrary-precision integer
Bernstein operations. Terminal entropy tests use outward interval arithmetic.
The joint correction accepts a cell only through rational or integer
inequalities; floating point may guide subdivision but does not certify a
sign.

The upper calculation checks integer conservation identities, speaker
locality, and terminal supports. Input-wise channel sums are enclosed by
outward IEEE-754 binary64 arithmetic with a specified pairwise reduction
tree. Basic operations and \texttt{nextafter} are assumed correctly rounded,
and fast-math reassociation is disabled. Logarithms are enclosed by range
reduction and an explicit finite atanh series with a proved remainder. The
final scalar transformation is also evaluated outward.

The classification computations are finite exact enumerations over Boolean
matrices, rectangles, and protocol trees. They do not depend on the numerical
lower- or upper-bound data.

The deposit is a computer-assisted certificate package, not a proof-assistant
formalization. The analytical reductions in Appendices~\ref{sec:lower} and
\ref{sec:upper} explain why the verified finite inequalities imply the
claimed information bounds. Numerical search may suggest coefficients or
protocols, but no optimizer-convergence assertion is used in a theorem.

The reproduction procedure specified here covers the lower- and upper-bound
certificates in Theorem~\ref{thm:main} and the classification in
Theorem~\ref{thm:classification}. The product-prior estimate in
Appendix~\ref{app:product-prior} is proved by the explicit finite
inequalities displayed there and does not depend on the multi-million-node upper certificate.

\section{Combinatorial witnesses for the \texorpdfstring{$3\times3$}{3x3} classification}
\label{app:classification}
This appendix makes the finite classification in
Theorem~\ref{thm:classification} independently checkable; it does not
use any of the numerical information-complexity certificates.

\subsection{Counting and testing the classes}
The $36$ row/column classes can be enumerated by selecting the
lexicographically least element of the orbit of each nine-bit matrix
under $S_3\times S_3$.  Their number can also be found without
enumeration: the number of matrices fixed by a pair of permutations is
$2^c$, where $c$ is the number of cycles of its action on the nine
cells.  The six row/column cycle-type pairs contribute
\[
\begin{array}{c|rrrrrr}
\text{types}&(1^3,1^3)&(1^3,21)&(1^3,3)&(21,21)&(21,3)&(3,3)\\
\hline
\text{number of pairs}&1&6&4&9&12&4\\
\text{fixed matrices per pair}&512&64&8&32&4&8
\end{array}
\]
respectively; their sum divided by $36$ is $36$.  Explicitly,
representatives of all row/column classes, with the maximum
rectangle-separated set size $k$, minimum leaf count $L$, and minimum
binary depth $D$, are listed in the following table.  Each entry
\texttt{abc/def/ghi} specifies rows in order.

\begin{center}
\footnotesize
\begin{tabular}{@{}lrrr@{\qquad}lrrr@{}}
\toprule
Representative & $k$ & $L$ & $D$ & Representative & $k$ & $L$ & $D$\\
\midrule
\texttt{000/000/000} & 1 & 1 & 0 & \texttt{000/011/111} & 4 & 4 & 2\\
\texttt{000/000/001} & 3 & 3 & 2 & \texttt{001/001/111} & 3 & 3 & 2\\
\texttt{000/000/011} & 3 & 3 & 2 & \texttt{001/010/111} & 5 & 5 & 3\\
\texttt{000/001/001} & 3 & 3 & 2 & \texttt{001/011/011} & 4 & 4 & 2\\
\texttt{000/001/010} & 4 & 5 & 3 & \texttt{001/011/101} & 5 & 5 & 3\\
\texttt{000/000/111} & 2 & 2 & 1 & \texttt{001/011/110} & 6 & 6 & 3\\
\texttt{000/001/011} & 5 & 5 & 3 & \texttt{001/110/110} & 4 & 4 & 2\\
\texttt{000/001/110} & 4 & 4 & 2 & \texttt{000/111/111} & 2 & 2 & 1\\
\texttt{001/001/001} & 2 & 2 & 1 & \texttt{001/011/111} & 5 & 5 & 3\\
\texttt{001/001/010} & 4 & 4 & 2 & \texttt{001/110/111} & 4 & 4 & 2\\
\texttt{001/010/100} & 6 & 6 & 3 & \texttt{011/011/011} & 2 & 2 & 1\\
\texttt{000/001/111} & 4 & 4 & 2 & \texttt{011/011/101} & 4 & 4 & 2\\
\texttt{000/011/011} & 3 & 3 & 2 & \texttt{011/101/110} & 6 & 6 & 3\\
\texttt{000/011/101} & 5 & 5 & 3 & \texttt{001/111/111} & 3 & 3 & 2\\
\texttt{001/001/011} & 4 & 4 & 2 & \texttt{011/011/111} & 3 & 3 & 2\\
\texttt{001/001/110} & 4 & 4 & 2 & \texttt{011/101/111} & 4 & 5 & 3\\
\texttt{001/010/011} & 5 & 5 & 3 & \texttt{011/111/111} & 3 & 3 & 2\\
\texttt{001/010/101} & 6 & 6 & 3 & \texttt{111/111/111} & 1 & 1 & 0\\

\bottomrule
\end{tabular}
\end{center}

For checking the witness sets, two cells $(i,j)$ and $(i',j')$
can belong to a common monochromatic rectangle exactly when all
entries of the rectangular hull
$\{i,i'\}\times\{j,j'\}$ have the same value. Consequently, a
rectangle-separated set is a set of cells for which no pair satisfies
this test. Checking all subsets of the nine cells finds its maximum size.

For protocol leaf counts, use the following recursion. For nonempty
row and column sets $A,B$, put $\ell_f(A,B)=1$ if $f$ is constant
on $A\times B$. Otherwise,
\begin{equation}
\label{eq:classification-recursion}
\ell_f(A,B)=\min\left\{
\begin{aligned}
&\ell_f(A',B)+\ell_f(A\setminus A',B)
  &&(\emptyset\ne A'\subsetneq A),\\
&\ell_f(A,B')+\ell_f(A,B\setminus B')
  &&(\emptyset\ne B'\subsetneq B).
\end{aligned}
\right.
\end{equation}
Each recursion step is precisely a message that partitions one
party's current possible input values into two sets. The corresponding
minimum-depth recursion uses terminal value zero and $1+\max$ in
place of the sum. There are
only $7\cdot7=49$ nonempty row/column subrectangles to consider
for each matrix.

\subsection{Explicit rectangle-separated sets and protocol trees}
For compactness, a cell is denoted by its two one-based indices
($ij$ means $(i,j)$). The following are explicit witness sets
for the thirteen families of Table~\ref{tab:classification};
rows and columns of any other member are obtained by the stated
symmetries.
\begin{center}
\small
\begin{tabular}{@{}ll@{}}
\toprule
Representative & Rectangle-separated input set $S$\\
\midrule
\texttt{000/000/000} & $\{11\}$\\
\texttt{000/000/001} & $\{13,\;31,\;33\}$\\
\texttt{000/000/011} & $\{12,\;31,\;32\}$\\
\texttt{000/001/010} & $\{12,\;23,\;31,\;32\}$\\
\texttt{000/000/111} & $\{11,\;31\}$\\
\texttt{000/001/011} & $\{13,\;22,\;23,\;31,\;32\}$\\
\texttt{000/001/110} & $\{11,\;23,\;31,\;33\}$\\
\texttt{001/010/100} & $\{11,\;13,\;22,\;23,\;31,\;32\}$\\
\texttt{000/001/111} & $\{13,\;21,\;23,\;31\}$\\
\texttt{000/011/011} & $\{12,\;21,\;22\}$\\
\texttt{000/011/101} & $\{13,\;21,\;22,\;31,\;32\}$\\
\texttt{001/001/110} & $\{11,\;13,\;31,\;33\}$\\
\texttt{001/010/101} & $\{11,\;13,\;22,\;23,\;31,\;32\}$\\

\bottomrule
\end{tabular}
\end{center}

Here is a compact encoding of verifying deterministic protocol trees.
A node \texttt{A12?(T):(U)} is an Alice message testing whether her
input lies in $\{1,2\}$, continuing with $T$ on yes and $U$ on no;
\texttt{B12?} has the analogous meaning for Bob. A terminal symbol
\texttt{0[12,123]} denotes the leaf rectangle
$\{1,2\}\times\{1,2,3\}$, with output zero. Tests always apply to
the input values still possible after preceding messages. All listed
trees have the leaf counts in Table~\ref{tab:classification}, and
every terminal rectangle is monochromatic.

\begin{lstlisting}[basicstyle=\ttfamily\scriptsize,breaklines=true,columns=fullflexible]
000/000/000 : 0[123,123]
000/000/001 : A12?(0[12,123]):(B12?(0[3,12]):(1[3,3]))
000/000/011 : A12?(0[12,123]):(B1?(0[3,1]):(1[3,23]))
000/001/010 : A1?(0[1,123]):(A2?(B12?(0[2,12]):(1[2,3])):(B2?(1[3,2]):(0[3,13])))
000/000/111 : A12?(0[12,123]):(1[3,123])
000/001/011 : A1?(0[1,123]):(A2?(B12?(0[2,12]):(1[2,3])):(B1?(0[3,1]):(1[3,23])))
000/001/110 : B12?(A12?(0[12,12]):(1[3,12])):(A2?(1[2,3]):(0[13,3]))
001/010/100 : A1?(B12?(0[1,12]):(1[1,3])):(A2?(B2?(1[2,2]):(0[2,13])):(B1?(1[3,1]):(0[3,23])))
000/001/111 : A2?(B12?(0[2,12]):(1[2,3])):(A1?(0[1,123]):(1[3,123]))
000/011/011 : A1?(0[1,123]):(B1?(0[23,1]):(1[23,23]))
000/011/101 : A1?(0[1,123]):(A2?(B1?(0[2,1]):(1[2,23])):(B2?(0[3,2]):(1[3,13])))
001/001/110 : A12?(B12?(0[12,12]):(1[12,3])):(B12?(1[3,12]):(0[3,3]))
001/010/101 : A1?(B12?(0[1,12]):(1[1,3])):(A2?(B2?(1[2,2]):(0[2,13])):(B2?(0[3,2]):(1[3,13])))
\end{lstlisting}

The deposited classification module contains a standard-library Python
enumeration and an independent audit of these witnesses.\zenodopath{classification/} These checks are independent of the large information-complexity certificates. No
statement about the exact randomized value for the exceptional
matrix relies on the enumeration: this finite classification shows
only that all other $3\times3$ classes have matching elementary
upper and lower bounds.

\bibliographystyle{alpha}
\bibliography{ic_ovr}
\end{document}